\documentclass[a4paper,fleqn,usenatbib]{mnras}

\usepackage{newtxtext}

\usepackage[T1]{fontenc}
\usepackage{ae,aecompl}

\usepackage{natbib,twoopt}
\usepackage{siunitx}
\usepackage[fleqn]{amsmath}
\usepackage{graphicx}
\usepackage{xcolor}
\usepackage{colortbl}
\usepackage{multirow}
\usepackage{scalefnt}

\usepackage{scalerel}
\usepackage{tikz}
\usetikzlibrary{svg.path}

\definecolor{orcidlogocol}{HTML}{A6CE39}
\tikzset{
  orcidlogo/.pic={
    \fill[orcidlogocol] svg{M256,128c0,70.7-57.3,128-128,128C57.3,256,0,198.7,0,128C0,57.3,57.3,0,128,0C198.7,0,256,57.3,256,128z};
    \fill[white] svg{M86.3,186.2H70.9V79.1h15.4v48.4V186.2z}
                 svg{M108.9,79.1h41.6c39.6,0,57,28.3,57,53.6c0,27.5-21.5,53.6-56.8,53.6h-41.8V79.1z M124.3,172.4h24.5c34.9,0,42.9-26.5,42.9-39.7c0-21.5-13.7-39.7-43.7-39.7h-23.7V172.4z}
                 svg{M88.7,56.8c0,5.5-4.5,10.1-10.1,10.1c-5.6,0-10.1-4.6-10.1-10.1c0-5.6,4.5-10.1,10.1-10.1C84.2,46.7,88.7,51.3,88.7,56.8z};
  }
}

\newcommand\orcidicon[1]{\href{https://orcid.org/#1}{\mbox{\scalerel*{
\begin{tikzpicture}[yscale=-1,transform shape]
\pic{orcidlogo};
\end{tikzpicture}
}{|}}}}

\usepackage{hyperref} 

\title[Co-Orbital Exomoons]{The Dynamics of Co-orbital Giant Exomoons - Applications for the Kepler-1625 b and Kepler-1708 b Satellite Systems}

\author[R. A. Moraes, G. Borderes-Motta, O. C. Winter and D. C. Mour\~{a}o]{%
  R. A. Moraes$^{1}$\thanks{E-mail: ricardo.moraes@unesp.br (RAM)}\orcidicon{0000-0002-4013-8878}\,
  G. Borderes-Motta$^{2,3}$\thanks{gabriel.borderes@uc3m.es (GBM)}\orcidicon{0000-0002-4680-8414}\, O. C. Winter$^{1}$\thanks{othon.winter@unesp.br (OCW)}\orcidicon{0000-0002-4901-3289}\ and D. C. Mour\~{a}o$^{1}$\thanks{daniela.mourao@unesp.br (DCM)}\orcidicon{0000-0001-9555-8143} \\
  $^{1}$ UNESP, S\~{a}o Paulo State University - Grupo de Din\^{a}mica Orbital \& Planetologia, Guaratinguet\'{a}, CEP 12.516-410, S\~{a}o Paulo, Brazil\\
  $^{2}$ UC3M, Univ. Carlos III de Madrid - Bioengineering and Aerospace Engineering Department, Legan\'{e}s, 28911, Madrid, Spain \\
  $^{3}$ IRF, Swedish Institute of Space Physics, Kiruna, 98128, Norrbotten, Sweden}
\date{Accepted XXX. Received YYY; in original form ZZZ}

\pubyear{2022}

\begin{document}
\label{firstpage}
\pagerange{\pageref{firstpage}--\pageref{lastpage}}
\maketitle


\begin{abstract}
Exomoons are a missing piece of exoplanetary science. Recently, two promising candidates were proposed, Kepler-1625 b-I and Kepler-1708 b-I. While the latter still lacks a dynamical analysis of its stability, Kepler-1625 b-I has already been the subject of several studies regarding its stability and origin. Moreover, previous works have shown that this satellite system could harbour at least two stable massive moons. Motivated by these results, we explored the stability of co-orbital exomoons using the candidates Kepler-1625 b-I and Kepler-1708 b-I as case studies. To do so, we performed numerical simulations of systems composed of the star, planet, and the co-orbital pair formed by the proposed candidates and another massive body. For the additional satellite, we varied its mass and size from a Mars-like to the case where both satellites have the same physical characteristics. We investigated the co-orbital region around the Lagrangian equilibrium point $L_4$ of the system, setting the orbital separation between the satellites from $\theta_{min} = 30^{\circ}$ to $\theta_{max} = 90^{\circ}$. Our results show that stability islands are possible in the co-orbital region of Kepler-1708 b-I as a function of the co-orbital companion's mass and angular separation. Also, we identified that resonances of librational frequencies, especially the 2:1 resonance, can constrain the mass of the co-orbital companion. On the other hand, we found that the proximity between the host planet and the star makes the co-orbital region around Kepler-1625 b-I unstable for a massive companion. Finally, we provide TTV profiles for a planet orbited by co-orbital exomoons.
\end{abstract}


\section{Introduction}
\label{sone}
To date, more than $5,000$ extrasolar planets, exoplanets, have been discovered. This increasing population of bodies presents an outstanding diversity of sizes, masses, and orbital characteristics. Based on the planets of our Solar System, one would expect an abundant population of natural satellites around exoplanets, the exomoons.

Exomoons are yet to be confirmed, but several candidates are reported in the literature \citep{Bennett-etal-2014, Ben-Jaffel-Ballester-2014, Lewis-etal-2015, Hippke-2015, Teachey-etal-2018, Heller-etal-2019, Oza-etal-2019, Fox-Wiegert-2021, Kipping-etal-2022}. However, these candidates should be regarded with caution. Some of these satellites are either dynamically unlikely, as is the case for the satellites proposed by \citet{Ben-Jaffel-Ballester-2014}, which would be orbiting outside the Hill sphere of their host planet, or simply false positives, for example, the six candidates proposed by \citet{Fox-Wiegert-2021} that were refuted from both an observational perspective \citep{Kipping-2020} and a stability perspective \citep{Quarles-etal-2020}. In addition, there are satellites proposed based only on indirect effects detected on the host planet \citep{Oza-etal-2019}. Among these candidates, the two most promising are Kepler-1625 b-I \citep{Teachey-etal-2018} and Kepler-1708 b-I \citep{Kipping-etal-2022}. Both bodies are predicted to be planet-like satellites and were proposed based on transit light curves from the \textit{Kepler Space Telescope} (\textit{Kepler}). 

The signs of Kepler-1625 b-I were first seen in three transit light curves from \textit{Kepler} \citep{Teachey-etal-2018} and subsequently identified in data from the \textit{Hubble Space Telescope} (\textit{HST}) \citep{Teachey-Kipping-2018}. However, the only transit from \textit{HST} did not provide enough information to settle the discussion. Moreover, further analysis of the transits found evidence both in favor \citep{Teachey-etal-2020} and against \citep{Rodenbeck-etal-2018, Heller-etal-2019, Kreidberg-etal-2019} the moon hypothesis, leaving Kepler-1625 b-I as a candidate.

Kepler-1708 b-I was the only emerging exomoon candidate from a survey of 70 transiting cool giant exoplanets compiled by \citet{Kipping-etal-2022} using \textit{Kepler}'s archive. Although there is only one transit exhibiting a faint signal supporting the newest candidate, this evidence checked all the criteria applied by the authors. Recently, \citet{Cassese-Kipping-2022} showed that Kepler-1708 b-I is unlikely to be detected by \textit{HST}, which leaves its status unresolved. Further analysis of the transit presented by \citet{Kipping-etal-2022} and more data are needed to validate or refute the candidate.

The lack of confirmed exomoons is due to various reasons, mainly technology limitations. The space telescopes \textit{Kepler} and \textit{CoRot} carried expectations on the potential detection of exomoons \citep{Szabo-etal-2006, Simon-etal-2007, Kipping-etal-2009}, but only a few candidates could be proposed using data from these facilities. The \textit{HST} also could give us hints about exomoons. However, only one transit light curve of \textit{HST} presented signs of natural satellites \citep{Teachey-Kipping-2018}. 

In the near future, a new generation of space telescopes is expected to expand the horizon of possibilities for exomoon detection. The recently deployed \textit{James Webb Space Telescope} (\textit{JWST}) could provide transit light curves to allow the confirmation of the exomoon candidates \citep{Kipping-etal-2022} in addition to detect new ones \citep{Limbach-etal-2021}. The 2026 \textit{PLATO} mission will search for exoplanets around stars brighter than the ones observed by \textit{Kepler} \citep{Rauer-etal-2014}, thus also increasing the possibility of finding exomoons around close-in giant planets \citep{Heller-book-2018, Hippke-Heller-2022}. 

While waiting for the new telescopes, astronomers have focused on developing techniques to search for signs of exomoons on the available data. Most of the efforts are directed towards the analysis of planetary transits, such that the exomoon's transit or its indirect effects on the planet's transit can be identified \citep{Sartoretti-Schneider-1999, Simon-etal-2007, Kipping-2009a, Kipping-2009b, Heller-2014, Kipping-2021, Teachey-2021}. In addition, \citet{Hippke-Heller-2022} released \texttt{PANDORA}, the first publicly available transit-fitting software to search for transits of exomoons, which aims to increase the number of scientists looking for exomoons. 

Exomoons can also be addressed from a theoretical point of view. In this way, it is possible to study the stability of exomoons, thus constraining the number, mass, and location of satellites around exoplanets.

Due to observation bias, most of the detected exoplanets are giant planets orbiting close to their host star. Naturally, the first studies about the stability of exomoons were focused on satellites around these planets. \citet{Barnes-Obrien-2002} applied tidal theory and numerical simulations to constrain the mass of satellites around close-in giant planets that would survive tidal migration and be long-term stable. Their results predicted that Earth-like moons could be stable around Jovian planets depending on the star's mass and the star-planet separation. However, \citet{Barnes-Obrien-2002} considered the planet's inner properties and rotation to be stable for billions of years, which does not correspond to reality. Over long periods of time, planets might shrink \citep{Fortney-etal-2007}, while their interior cools and hardens \citep{Guenel-etal-2014}. \citet{Alvarado-Montes-etal-2017}, \citet{Sucerquia-etal-2019} and \citet{Sucerquia-etal-2020} described the tidal migration of exomoons using more robust models. \citet{Alvarado-Montes-etal-2017}, for example, focused on the radius contraction and internal structure evolution of the planets to constrain the tidal migration of satellites around close-in giant planets. The authors proposed that exomoons exposed to tides would have three fates: 1) fall into the planet; 2) orbital detachment, being ejected from the planet's Hill sphere; 3) migration to a place of stability, in a quasi-stationary orbit. They found that the inward migration leading to fate 1) is suppressed, such as the exomoons will migrate towards an asymptotic maximum distance leading to fate 3) where the satellites will be stable for long periods of time. Later, \citet{Sucerquia-etal-2020} named this distance as \textit{satellite tidal orbital parking}.

\citet{Domingos-etal-2006} numerically studied the stability of prograde and retrograde orbits of satellites around giant planets. The authors were able to derive analytical expressions for the critical semimajor axis for the stability of exomoons in both prograde and retrograde motion. The estimations presented by \citet{Domingos-etal-2006} were recently updated by \citet{Rosario-Franco-etal-2020}. From an analytic perspective, \citet{Donnison-2010} applied the Hill stability criteria \citep{Murray-1999} to place limits on the critical separation between planets and stable satellites for various planet-satellite mass ratios. On the other hand, \citet{Namouni-2010} showed that Galilean-like satellites are unlikely to survive the inward migration of their host planet. During an inwards migration, the planet's Hill radius shrinks, such as the orbits of hypothetical satellites would become more unstable, leading to the ejection of the moons. The results of \citet{Namouni-2010} could be related to the lack of satellites detected so far around close-in giant planets.

After the announcement of Kepler-1625 b-I, new models for the stability and formation of exomoons were designed specifically for this candidate. Studies have shown that the origins of this candidate could be explained by both capture \citep{Heller-2018, Hamers-Zwart-2018, Hansen-2019} and in-situ formation \citep{Moraes-Vieira-Neto-2020}. However, as these models aimed to reproduce the characteristics of Kepler-1625 b-I as reported by \citet{Teachey-etal-2018} and \citet{Teachey-Kipping-2018}, their applicability to other satellite systems cannot be assured \citep{Kipping-etal-2022}.

Given the size, mass, and orbital separation of Kepler-1625 b-I \citep{Teachey-etal-2018,Teachey-Kipping-2018}, it is expected that this satellite candidate underwent an intense phase of tidal migration after its formation or capture. Assuming that the host planet is a Jupiter-like body, several studies showed that a Neptune-like satellite would survive the tidal interactions with the host planet and be stable for long periods of time \citep{Rosario-Franco-etal-2020, Tokadjian-Piro-2020, Quarles-Rosario-Franco-2020}. In addition to the satellite candidate's stability, \citet{Kollmeier-Raymond-2019, Rosario-Franco-etal-2020} showed that theoretical submoons could also be possible around Kepler-1625 b-I. Moreover, \citet{Moraes-etal-2022} showed that an extra Earth-like satellite could be stable in the Kepler-1625 b satellite system. The authors explored the regions inside the predicted orbit of Kepler-1625 b-I and found that planetary and satellite tides could stabilize inner satellites, such as these bodies will not fall into the planet or migrate outwards and collide with the satellite candidate. Also, the authors pointed out that the formation of mean motion resonances between the satellites is a dynamical mechanism that secures the stability of both satellites, even when one of the moons has an eccentric orbit. 

As it was only recently announced, there are only a few theoretical studies assessing the origins and stability of the exomoon candidate Kepler-1708 b-I. \citet{Kipping-etal-2022} studied the post-formation tidal evolution of the candidate, assuming that the satellite was formed in-situ at twice the Roche limit. The authors found that once the moon is initially beyond the corotation radius of the system with low eccentricity, the satellite migrates outwards to its proposed position. The authors pointed out that this result does not contribute to determining a possible formation scenario for the candidate, owing to the fact that any model that forms a massive satellite in a tight configuration with the planet could reproduce the tidal outwards migration they presented.

As shown by \citet{Moraes-etal-2022}, more than one massive moon might be stable in the Kepler-1625 b satellite system. Here, we investigate the possibility of extra satellites being stable in a co-orbital configuration with Kepler-1625 b-I and Kepler-1708 b-I. In our Solar System, several satellites can be found exhibiting co-orbital motion, for example, the Saturnian satellite Tethys and its co-orbital companions Telesto and Calypso, and the classic example of satellites in co-orbital motion, Janus and Epimetheus also in the Saturn system.

Since Kepler-1625 b-I is predicted to be a Neptune-like satellite and Kepler-1708 b-I is expected to be a Super-Earth-like body, we aim to search for stable co-orbital companions that are also planet-like bodies. As shown by \citet{Gascheau-1843} and \citet{Routh-1874}, for the general three-body problem the Lagrangian points $L_4$ and $L_5$ can be linearly stable depending on the mass of the bodies. 

Before going any further describing our approach to the proposed problem, we must answer the following question: Is it possible to have co-orbital systems composed of planet-like bodies?

As mentioned before, we have co-orbital satellites in our Solar System. However, they are usually formed by two bodies with not comparable masses, with the Janus-Epimetheus system as a remarkable exception, around a planet that is significantly bigger. On the other hand, Kepler-1625 b-I and Kepler-1708 b-I are planet-like satellites, and we are proposing that these bodies have planet-like co-orbital companions. In this way, mechanisms that produce co-orbital planets could be applied to planet-like satellites.

Recently, \citet{Long-etal-2022} presented strong evidences that dust could be trapped around the Lagrangian points of a young planet candidate immersed on the LkCa 15 disk using data from the ALMA telescope. This potential discovery was already theoretically predicted in the literature. \citet{Montesinos-etal-2020} showed that up to cm-sized particles could be trapped around Lagrangian points in protoplanetary disks. The authors identified that the formation of local vortices at $L_4$ and $L_5$ in the early stages of planetary formation are responsible for this dust accumulation. If dust can agglomerate at the Lagrangian points of a giant planet, then the in-situ formation of Earth-size co-orbital companions might be possible due to the coagulation of these particles into a single massive rocky object \citep{Chiang-Lithwick-2005, Beauge-etal-2007, Lyra-etal-2009, Giuppone-etal-2012}. In addition, \citet{Laughlin-Chambers-2002} used hydrodynamic simulations to show that a pair of Jupiter-like co-orbital planets could be formed by accretion in a protoplanetary disk, and this 1:1 resonance configuration would be sustainable even after the inwards migration of the planets. Other methods such as pull-down capture of a companion into co-orbital orbit, formation due to direct collision \citep{Chiang-Lithwick-2005}, convergent migration of multiple protoplanets \citep{Thommes-2005, Cresswell-Nelson-2006} and gravitational scattering of planetesimals by a protoplanet \citep{Kortenkamp-2005} have been proposed as alternative methods to the in-situ formation for the origins of co-orbital planets.

For the formation of co-orbital massive satellites, \citet{Moraes-Vieira-Neto-2020} showed that in-situ in a massive-solid-enhanced circumplanetary disk could explain the origins of Kepler-1625 b-I. Also, the authors showed that other satellites could form as well. This formation mechanism is compatible with the formation of co-orbital planets if dust could accumulate at the Lagrangian point of the satellite. On the other hand, if Kepler-1625 b-I was captured in a compact configuration, this object could capture pre-existing surviving satellites into co-orbital orbits during its outwards migration phase. \citet{Heller-2018} proposed that Kepler-1625 b-I was part of a binary planetary system before being captured and that the other part of the binary was ejected during the capture phase. It is tempting to suppose that if the binary was captured intact, a co-orbital satellite system could survive. However, the author did not explore this hypothesis, and the successful capture of the complete binary seems unlikely. Another possibility is that the planet-like satellite has its own submoons that ended up being detached from the satellite and later captured into a co-orbital configuration. Even though this is a valid hypothesis, given that submoons are stable only to one-third of the satellite's Hill radius \citep{Rosario-Franco-etal-2020}, here we will be simulating co-orbital companions that are much more massive than the mass limit for submoons proposed by \citet{Kollmeier-Raymond-2019} and \citet{Rosario-Franco-etal-2020}.

In order to study the stability of massive co-orbital satellites in the Kepler-1625 b and Kepler-1708 b systems, we will consider two scenarios. First, the simulated systems will be composed of the planet, the satellite candidate, and the co-orbital companion, neglecting the presence of the star. This assumption considers that Kepler-1625 b-I and Kepler-1708 b-I are both well inside the stability limit for exomoon's stability \citep{Domingos-etal-2006, Rosario-Franco-etal-2020}. In this case, we will be working with a general three-body problem. Subsequently, we will add the star of each system to the simulations. As we will see, co-orbital architectures are sensitive to perturbations, and initial conditions, such as even weaker gravitational effects, could break a once stable configuration, thus justifying the presence of the star. In addition, we want to explore the detectability of systems with co-orbital satellites. In this way, we will study the influences of co-orbital satellites over the planet's Transit Timing Variations (TTVs).

The stability of the co-orbital region of Kepler-1625 b-I and Kepler-1708 b-I will be studied considering co-orbital companions with different: masses, sizes, and initial angular positions. The radius of the bodies is taken into account only to allow collisions. Here we do not consider any non-gravitational effect.

This paper is organized as follows. In Sec. \ref{stwo}, we present a review of the physical and orbital characteristics of the bodies forming the Kepler-1625 and Kepler-1708 systems, the initial conditions explored for the extra satellite in each case, and our numerical tools. Then, in Secs. \ref{sthree} and \ref{sfour}, we discuss our results regarding the stability and amplitude of libration of co-orbital satellites, the role of resonances on the surviving of the satellites, how the star of each system influences the stability of the co-orbital region, and an analysis of the planet's TTVs for the cases where co-orbital satellites are possible. In Sec. \ref{conclusion} we summarize our results and draw our conclusions.


\section{Model}\label{stwo}
In this section, we describe some physical and orbital characteristics of the Kepler-1625 and Kepler-1708 systems, the different systems and initial conditions considered, and present the numerical methods used in this work.

\subsection{The Kepler-1625 system}

The Kepler-1625 system is composed of a G-type star with approximately $8.7\pm2.1$ Gyr \citep{Teachey-Kipping-2018}, which names the system. The star is located in the Cygnus constellation, and it is a solar-mass object ($M_{\star} \sim 1.079$ $M_{\odot}$) with radius $R_{\star} \sim 1.793$ $R_{\odot}$ \citep{Mathur-etal-2017}. 

To date, only one planet has been detected in the system, Kepler-1625 b. The planet was first detected via planetary transit in 2015 with data from \textit{Kepler} \citep{Mullally-etal-2015} and confirmed in 2016 \citep{Morton-etal-2016}. Kepler-1625 b has a predicted semimajor axis of $a_p \sim 0.87$ au \citep{Morton-etal-2016, Heller-2018} and its believed to have a coplanar and circular orbit \citep{Teachey-etal-2018}. The planet has a radius of $R_p = 1.18$ Jupiter's radius ($R_J$), however, its mass is still not well constrained. Recent photodynamical models showed a distribution of mass peaking at $M_p = 3$ Jupiter's masses ($M_J$) as the most likely value for the mass of Kepler-1625 b (Fig. 10 from \citet{Teachey-etal-2020}).

The exomoon candidate Kepler-1625 b-I was initially predicted to be a Neptune-like body, with a semimajor axis of $\sim 19.1$ $R_p$ and a circular inclined orbit \citep{Teachey-etal-2018}. The inclination found for the satellite depends on the detrending method used for transit reduction. \citet{Teachey-Kipping-2018} found that the candidate's inclination will be $42^{+15\circ}_{-18}$ for linear detrending, $49^{+21\circ}_{-22}$ for quadratic detrending, and $43^{+15\circ}_{-19}$ for exponential detrending. In the same work, the author refined the semimajor axis of the satellite, also varying from one data reduction to another, $45^{+10}_{-5}$, $36^{+10}_{-13}$, and $42^{+7}_{-4}$ $R_p$ for linear, quadratic, and exponential detrending, respectively. Because of these uncertainties, many theoretical studies adopted the canonical value of $40$ $R_p$ for the semimajor axis of the satellite \citep{Hamers-Zwart-2018, Moraes-Vieira-Neto-2020, Moraes-etal-2022, Sucerquia-etal-2022}, which agrees with the prediction given by \citet{Martin-etal-2019}. However, other authors used different values for the planet-satellite separation \citep{Tokadjian-Piro-2022a}.

\subsection{The Kepler-1708 system}

The Kepler-1708 system is formed by a star, a planet, and the recently proposed exomoon. The star is an F-type, Sun-like object located in the Cygnus constellation. The mass and radius of this body are $M_{\star} \sim 1.1$ $M_{\odot}$ and $R_{\star} \sim 1.1$ $R_{\odot}$, respectively \citep{Kipping-etal-2022}. The age of the system is estimated to be around $3.16$ Gyr.

The planet Kepler-1708 b was first detected in 2011 by the \textit{Kepler} mission, but only recently validated by \citet{Kipping-etal-2022}. The planet has an estimated semimajor axis of $1.64^{+0.10}_{-0.10}$ au and is classified as a cool-giant. The planet's eccentricity is not well determined, and only an upper limit can be found, $e_p < 0.4$. The same is true for the physical characteristics of the planet. Its mass has an upper bound of $4.6$ $M_J$ and a radius of $R_p \sim 0.89 $ $R_J$ \citep{Kipping-etal-2022}. Assuming that the planet has a density similar to Jupiter, \citet{Tokadjian-Piro-2022a} set the mass of Kepler-1708 b to be $M_p = 0.81$ $M_J$.
The exomoon candidate is predicted to be a Super-Earth-like body with radius $R_s = 2.61^{+0.42}_{-0.43}$ $R_{\oplus}$ and an upper limit for the mass of $M_s < 37$ $M_{\oplus}$ \citep{Kipping-etal-2022}. To find a better estimation for the satellite's mass, \citet{Tokadjian-Piro-2022a} set the density of this body to be approximately equal to Neptune's. In this way, one can find $M_s = 5$ $M_{\oplus}$. \citet{Kipping-etal-2022} also found estimations for the orbital radius, $R_s = 11.7^{+3.8}_{-2.2}$ planetary radius, and inclination, $9^{+38\circ}_{-45}$, of the candidate. Despite the uncertainties, if confirmed, the satellite is expected to have low inclination. The motion of Kepler-1708 b-I is predicted to be circular around the host planet.

In Tab \ref{tab:properties} we present the canonical values for the systems Kepler-1625 and Kepler-1708 adopted in this work.

\begin{table*}
\caption[Properties]{Canonical parameters adopted in this work for the systems Kepler-1625 and Kepler-1708. We consider the planets and satellites in circular and coplanar orbits, except when stated differently.}
\begin{tabular}{ccccccccc}\hline 
 System & $M_{\star}$	    & $R_{\star}$ 	& $M_p$       & $R_p$     & $a_p$    & $M_s$     & $R_s$     & $a_s$			\\
 &$M_{\odot}$    & $R_{\odot}$   & $M_{J}$   & $R_{J}$ & $ua$     & $M_{\oplus}$ & $R_{\oplus}$ & $ R_p$         	\\ \hline	
 Kepler-1625 & $1.079$        & $1.793$       & $3.0$      & $1.18$    & $0.87$   & $17.15$     & $3.865$     & $ 40.0$          \\ \hline
 Kepler-1708 & $1.1$        & $1.1$       & $0.81$      & $0.89$    & $1.64$   & $5.0$     & $2.61$     & $ 11.7$ \\ \hline
\end{tabular}
\label{tab:properties}
\end{table*}

\subsection{Initial Conditions}

\subsubsection{Models without the star - Local system}
\citet{Domingos-etal-2006} and \citet{Rosario-Franco-etal-2020} calculated the stability region for exomoons around a planet. The most conservative limit proposed in the aforementioned analysis points to exomoons in circular and coplanar orbits being stable if they are located inside $0.40$ Hill radius of the host planet. Neglecting the planet's eccentricity, the Hill radius of a planet can be written as,
\begin{align}\label{eq:hill}
    R_{H,p} = a_p\sqrt[3]{\dfrac{M_p}{3M_{\star}}}.
\end{align}
Using the canonical values for the systems Kepler-1625 and Kepler-1708 given in Tab. \ref{tab:properties}, from Eq. \ref{eq:hill}, one can see that exomoons Kepler-1625 b-I and Kepler-1708 b-I have, respectively, $a_s \sim 0.264$ $R_{H,p}$ and $a_s \sim 0.047$ $R_{H,p}$. In both cases, the proposed satellites are well inside the stability limit of $0.40$ $R_{H,p}$. Thus, these bodies are gravitationally influenced mainly by their parent planet, such as the star of each system should not play a significant role in their orbital evolution and could be neglected.

To model our systems, we consider a planetocentric coordinate system, with the planet as the central body. The satellite candidate, Kepler-1625 b-I or Kepler-1708 b-I, (hereafter "primary satellite"), and the extra satellite (hereafter "co-orbital companion") are initially in a co-orbital configuration, i.e., the satellites have the same semimajor axis and are only angularly separated.

\citet{Moraes-etal-2022} showed that the hypothetical Kepler-1625 b satellite system could be stable with two massive exomoons. The authors drew their conclusions after finding that an extra Earth-like satellite could survive in orbits internal to Kepler-1625 b-I. Here, we will extend this work investigating the possibility of co-orbital satellites in this system and in the Kepler-1708 system. To do so, we will explore a wide range of masses for the co-orbital companion since the stability in the three-body problem involving co-orbital bodies is very sensitive to the mass ratio between the co-orbital pair and the central body.

In our simulations, the mass and radius of the planets and the satellite candidates are taken from Tab. \ref{tab:properties}. In all cases, the satellites are in circular and coplanar orbits around the respective planet. The planet-satellite separation is also presented in  Tab. \ref{tab:properties}.

For the Kepler-1625 system, we consider $18$ different types of bodies as co-orbital companions, from Mars-sized to Neptune-sized. We varied the masses of these bodies from $M_2 = 0.107$ to $17.15$  $M_{\oplus}$, with the intermediate bodies having $M_2 = i$ $M_{\oplus}$, $i=1,\cdots, 16$. The radius of the satellites was interpolated using cubic splines taking the values of Mars, Earth, and Neptune as inputs.

Similarly, for the Kepler-1708 system, we have $12$ different types of co-orbital companions. The smaller and lighter body is a Mars-like companion, while for the other cases, we consider bodies with $M_2 = 0.5$ to $5$ $M_{\oplus}$, with $\Delta M_2 = 0.5$ $M_{\oplus}$ and radius interpolated as before.

\begin{figure}
\begin{center}
\includegraphics[height=0.8\columnwidth, width=0.8\columnwidth]{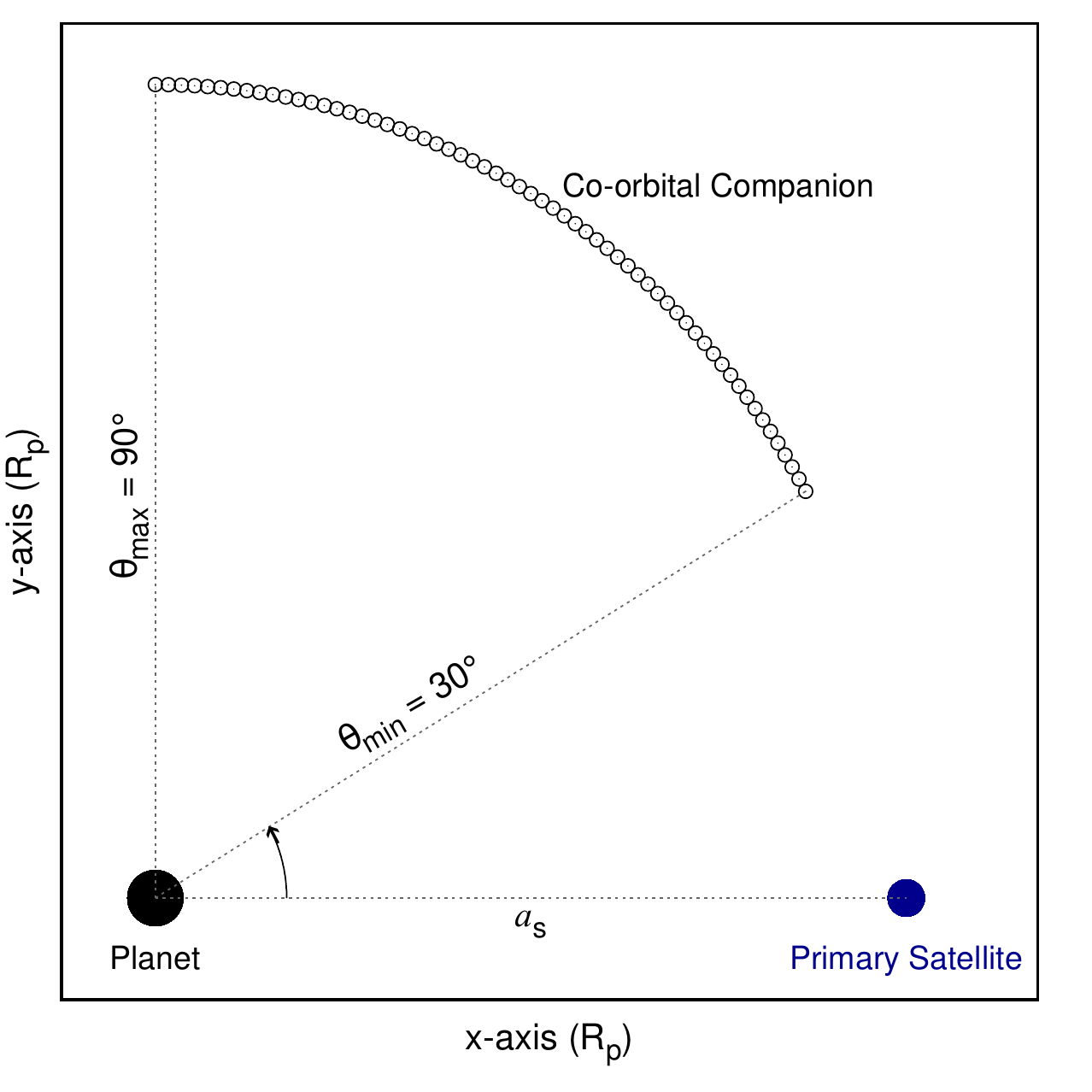}
\caption[Initial-setup]{Illustration of our system's initial set-up. The planet (black circle) is at the origin of the coordinate system, the primary satellite (blue circle) is placed at $a_s$ from the planet, and the co-orbital companion (open circles) is initially placed at $a_s$ from the planet with an angular separation $\theta$ measure from the horizontal line that connects the planet and the primary satellite anticlockwise, from $30^{\circ}$ to $90^{\circ}$.}
\label{fig:initial-setup}
\end{center}
\end{figure}

Once we set the characteristics of the co-orbital companion, we shall explore the initial angular position related to the primary satellite. Fig. \ref{fig:initial-setup} illustrates the initial set-up of our systems. We opted to vary the initial angular separation between the co-orbital satellites from $\theta_{min} = 30^{\circ}$ to $\theta_{max} = 90^{\circ}$, which means we will be exploring the surroundings of the Lagrangian equilibrium point $L_4$. In preliminary investigations, we found that for angular separations lesser than $30^{\circ}$, the co-orbital structure of the satellites is instantaneously destroyed. 

Because of the symmetry of the problem, the results and discussions presented in the following sections are the same for the respective regions around $L_5$. Thus, we will not analyze the case with the co-orbital companion near $L_5$.

We consider the satellites to be coplanar to the respective planet and initially in circular orbits. However, Kepler-1625 b-I is thought to be in an inclined configuration \citep{Teachey-etal-2018}, but since we are neglecting the presence of the star, we can assume that both satellites are in the same plane as the planet. 

\subsubsection{Models with the star - Complete system}
After our initial analysis, we will include the star of the systems and study its gravitation effects on the stability of the co-orbital architectures. For example, Kepler-1625 b has a semimajor axis smaller than $1$ au, and it is a $3$ Jupiter's masses planet. In this case, the gravitational interaction between the planet and the star could generate a non-negligible movement on the centre of mass of  the system, which ultimately translates as a wobble on the star. This change in the centre of mass of the system will induce additional movement on the planet and thus jeopardize the fate of co-orbital satellites. Also, by adding the star to our simulations we can access the effects of co-orbital moons on the TTVs of the planets.

The set-ups of the simulations with the star are the same as described before without the star. The planets are considered in circular orbits with the semimajor axis taken from Tab. \ref{tab:properties}.
Regarding the planet's inclination, Kepler-1708 b and its satellites are considered coplanar to the star. For the Kepler-1625 system, we study the case with the planet coplanar and with an inclination of $45^{\circ}$ relative to the star (other inclinations could be chosen, given the uncertainties on this parameter \citep{Teachey-etal-2018}).

\subsection{Numerical Tools}
For our study, we rely on numerical simulations to properly follow the time evolution of these systems since all the bodies involved gravitationally interact with each other.

Our numerical simulations were performed using the IAS15 integration scheme \citep{Rein-Spiegel-2015} implemented in the package POSIDONIUS \citep{Blanco-Cuaresma-Bolmont-2017b}. POSIDONIUS is often used in problems involving tides or other dissipative effects. However, because of familiarity with this numerical package, we found it convenient to make use of IAS15 written in RUST and opted simply to disable all the dissipative effects implemented in POSIDONIUS, computing only the gravitational interaction between the bodies. For comparison, we randomly chose some of our initial conditions and also simulated with the widely used REBOUND \citep{Rein-Liu-2012} package. The results produced by both packages showed good agreement with compatible computational times.

\section{Results for the local system: Planet-Satellite-Co-orbital companion}\label{sthree}

In this section, we present our results regarding the stability, shape, and amplitude of the co-orbital exomoons' orbits for the systems Kepler-1625 and Kepler-1708 considering a local system composed of the host planet, the satellite candidate, and the co-orbital companion.

\subsection{Stability}
Firstly, we present our results for the stability of the systems. We consider a system stable if both satellites are still co-orbitals at the end of the simulations. If the co-orbital configuration is destroyed, we label the system as unstable, regardless of the fate of the satellites after they leave their co-orbital architecture.

As the literature about the dynamics of two massive co-orbital bodies is limited, we will also use results from the co-orbital restricted three-body problem as a first approximation of our findings, such that a comparison between these results can be established.

\begin{figure*}
\begin{center}
\includegraphics[height=0.38\linewidth, width=0.49\linewidth]{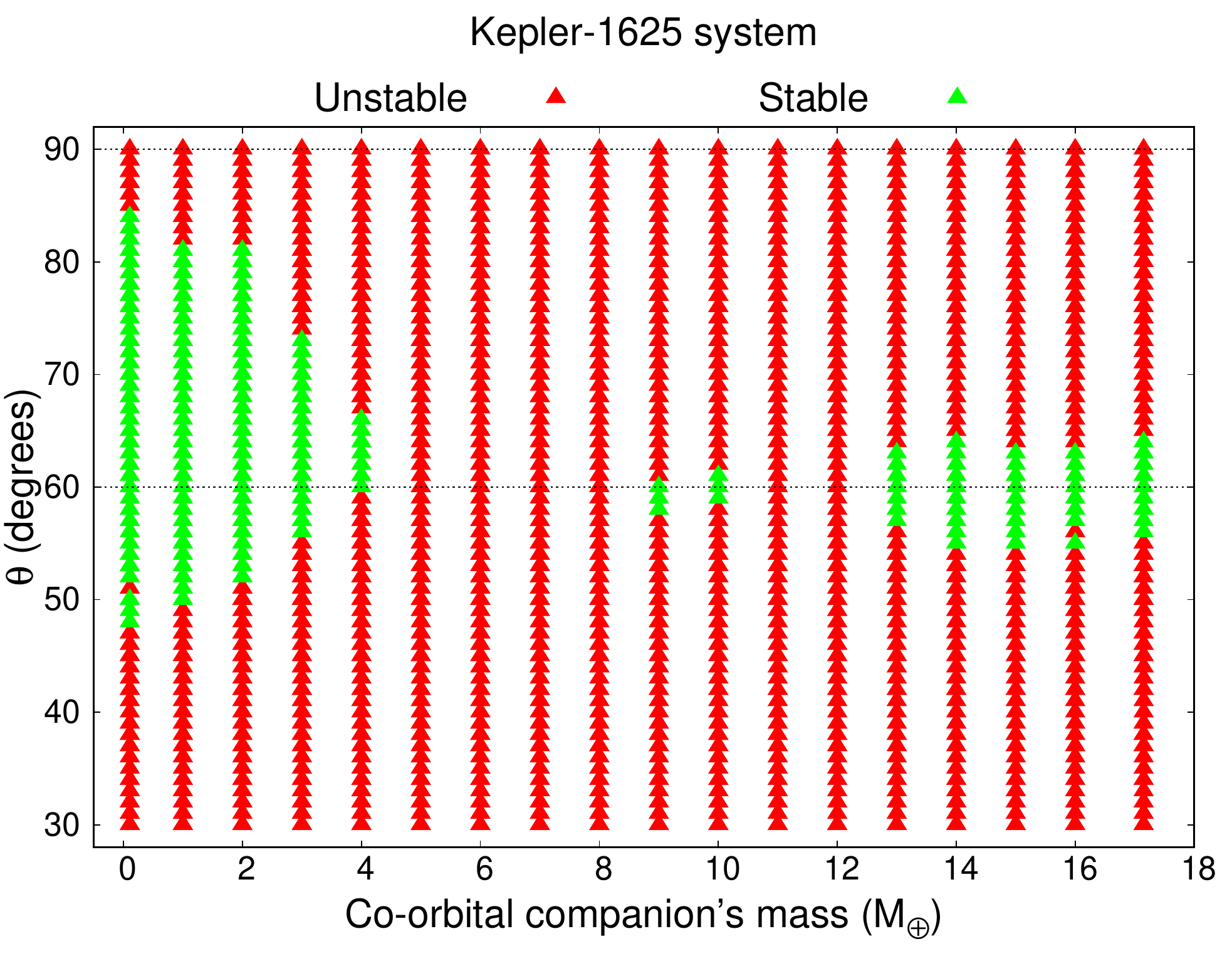}
\includegraphics[height=0.38\linewidth, width=0.49\linewidth]{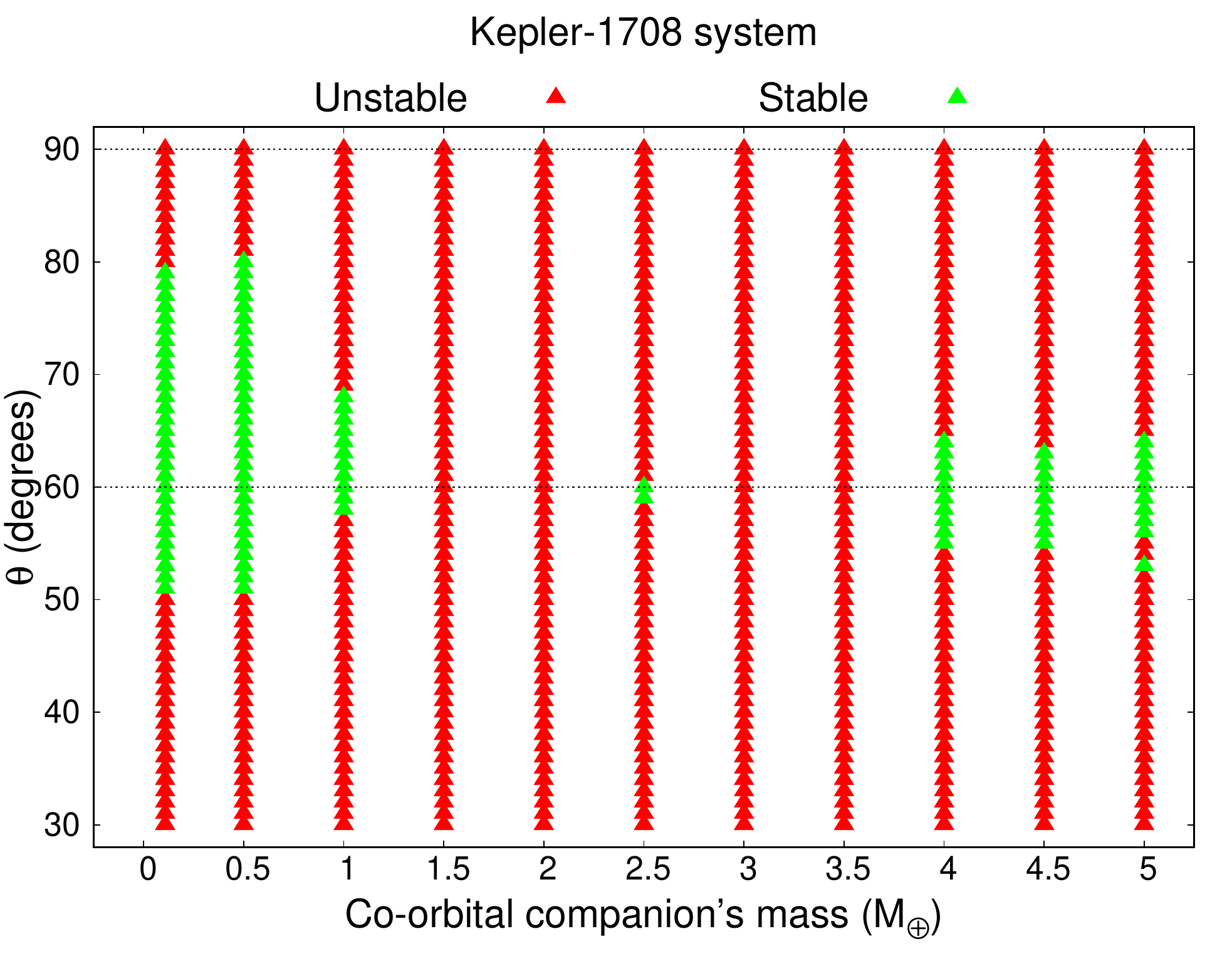}

\caption[Stability]{Grid with the initial conditions for the Kepler-1625 system (left panel) and Kepler-1708 system (right panel). In green, we have the conditions that end up being stable after $1$ Myr, and in red the unstable cases. The horizontal dotted lines mark initial conditions at $60^{\circ}$ and $90^{\circ}$.}
\label{fig:stability}
\end{center}
\end{figure*}

In Fig. \ref{fig:stability} we present our grid of initial conditions (mass of the co-orbital companion versus angular separation) for the system Kepler-1625 (left panel) and Kepler-1708 (right panel), respectively. In green, we have the initial conditions that became stable systems, and in red the unstable conditions. The simulations were carried out for $1$ Myr. As one can see, in both cases, the initial conditions for stable systems are around $L_4$ ($\theta = 60^{\circ}$), which is an equilibrium point in the restricted three-body problem. The stability of $L_4$ for the general three-body problem was already predicted, but not shown, by \citet{Erdi-Sandor-2005}, here we find that to be true for some cases. Also, there is a pronounced asymmetry for conditions around $L_4$, especially when the co-orbital companion is less massive. These asymmetries are due to the shape of the co-orbital companion's orbit, which is more elongated in the opposite direction of the primary satellite. This feature will be explored in detail when we address the shape of the satellites' orbit. 

For all the unstable systems, we found one of the following fates for the satellites: (i) collision between the satellites, or; (ii) collision between one satellite and the planet, or; (iii) ejection of one satellite. Thus, for the local systems we did not find in our simulations systems where both satellites survived after leaving their co-orbital configuration. 

For the Kepler-1625 system, we can see a correlation between the stability and mass of the co-orbital companion (left panel of Fig. \ref{fig:stability}). The entire co-orbital region is unstable when the extra satellite has $M_2 = 5 - 8$ $M_{\oplus}$ and $M_2 = 11 - 12$ $M_{\oplus}$, and not even the equilibrium point at $\theta = 60^{\circ}$ gave birth to stable systems. For $M_2 = 9 - 10$ $M_{\oplus}$, only a small region close to $L_4 $ is stable. On the other hand, for $M_2 \ge 13$ $M_{\oplus}$ the number of stable systems increases, which is counterintuitive. As the mass of the co-orbital companion increases, gravitational interactions between the satellites become stronger. In this way, one would expect the systems' stability to be compromised. However, we found the opposite. As we increased the mass of the co-orbital companions, more stable systems were found.

The same behaviour is seen for the system Kepler-1708 (right panel of Fig. \ref{fig:stability}). The unstable region as a function of the secondary satellite's mass extends from $M_2 = 1.5$ to $2.0$ $M_{\oplus}$ and from $M_2 = 3.0$ to $3.5$ $M_{\oplus}$. Same as before, in between these two unstable islands, there is a local stability close to $L_4$ for the systems with $M_2 = 2.5$ $M_{\oplus}$. Similar to the results for the system Kepler-1625, after the unstable valley, we found more stable conditions as the masses of the secondary satellites are increased.

The above-mentioned results suggest that the instability found for certain values of $M_2$ is caused by some dynamic effect that depends on the mass of the co-orbital companion. This effect will be explored in the following. 

\subsection{Resonances of Libration Frequency}
In the restricted three-body problem, the motion of a co-orbital particle about the $L_4$ of the system is composed by the superposition of two motions \citep{Murray-1999}. The first motion is a long-period motion of an epicentre librating about the $L_4$ of the system. Around this epicentre, the co-orbital satellite executes a short-period epicyclic motion, such as the final motion will be the summation of these two movements (Figs. 3.14 and 3.15 from \citet{Murray-1999}).

\begin{figure*}
\begin{center}
\includegraphics[height=0.75\columnwidth, width=\columnwidth]{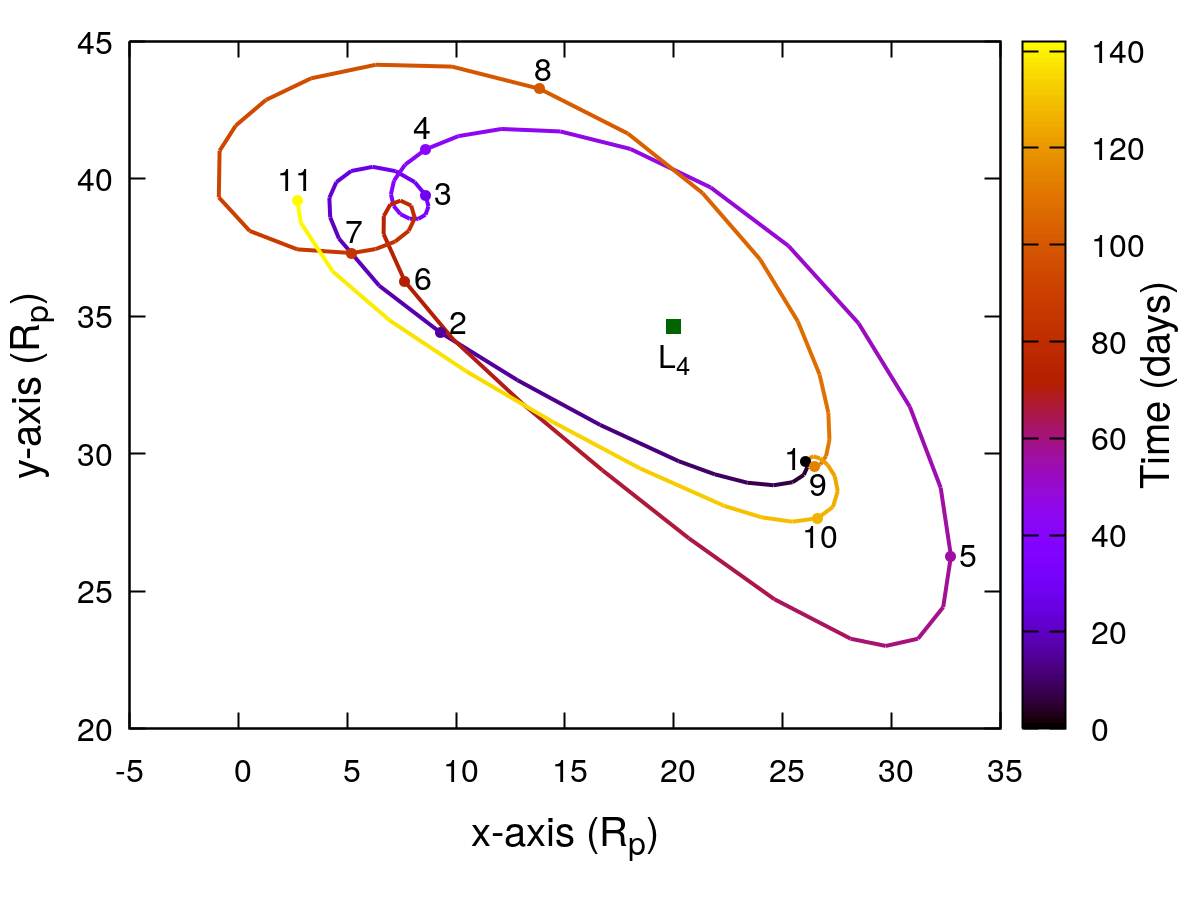}
\includegraphics[height=0.75\columnwidth, width=\columnwidth]{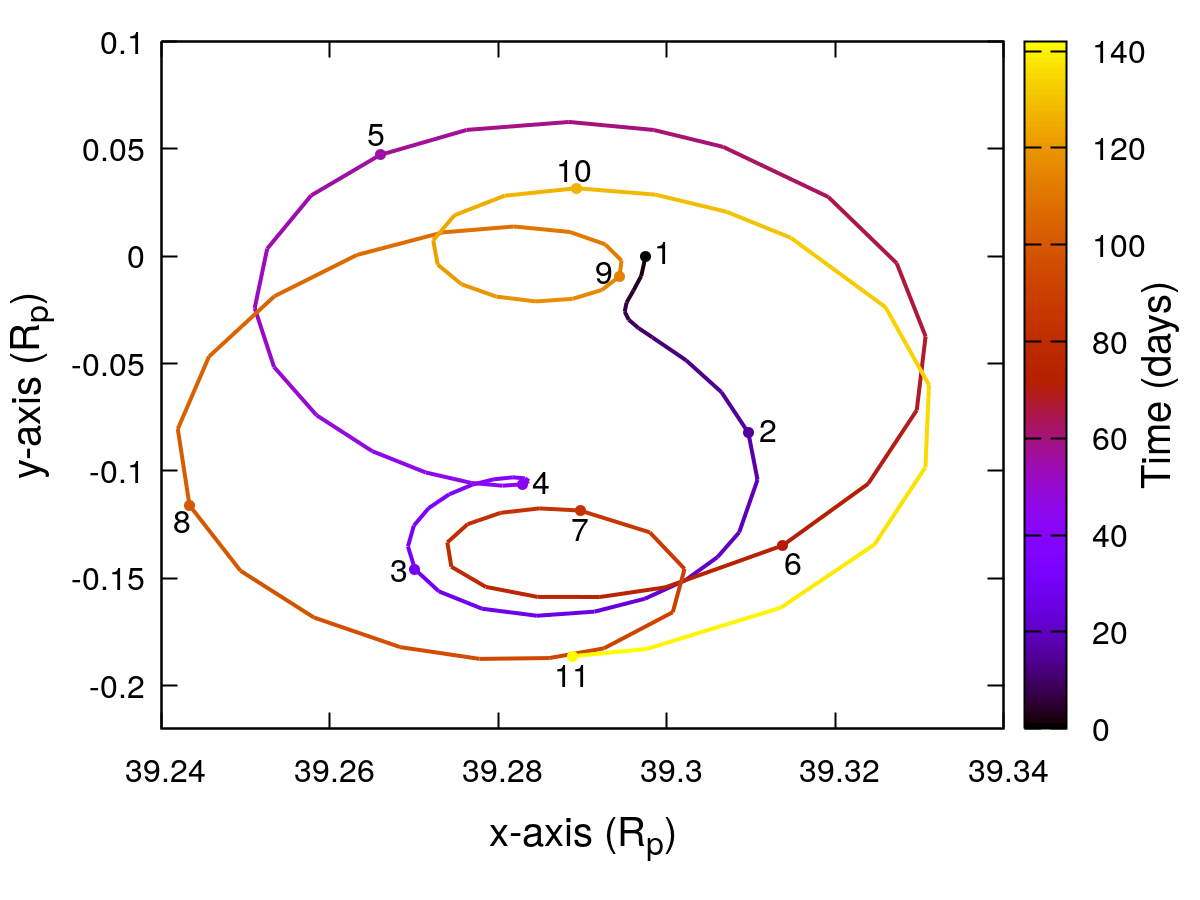}
\caption[Motions]{Motion of the co-orbital (left panel) and the primary satellite (right panel) in the Kepler-1625 system. The orbits of the satellites are depicted in the rotating frame for $142$ years. The co-orbital companion has $0.107$ $M_{\oplus}$ (Mars-size) and the satellites were initially $\theta = 48^{\circ}$ apart from each other. The colour bar indicates the time. The numbered points show the trajectory sequence of both satellites, such as corresponding positions have the same number.}
\label{fig:motion}
\end{center}
\end{figure*}

Although some similarities between the restricted and general three-body problems are expected, as the co-orbital satellites we are simulating are not particles, they will gravitationally affect the primary satellite, in which both bodies will librate with a particular frequency. To illustrate this behaviour, we show in Fig. \ref{fig:motion} the motion of the co-orbital (left panel) and the primary satellite (right panel) in the frame rotating with the initial circular frequency for $142$ years. For this example, we are considering the Kepler-1625 system, where the co-orbital companion is a Mars-size body and the satellites are initially $48^{\circ}$ apart from each other. The same pattern of motion was found for the satellites in the Kepler-1708 system.

As one can see, the motions of both satellites depicted in Fig. \ref{fig:motion} are similar to the motion of a particle about $L_4$ in the restricted three-body problem. The epicentre of the co-orbital companion is librating around $L_4$, while a short-period epicyclic pattern is observed in the loops of the satellite's trajectory. The same is true for the primary satellite, but in this case, the motion is performed near its initial position. The trajectory of both satellites in the rotating frame is tadpole-like, where the amplitude of the orbits is proportional to the perturbations felt by each satellite. 

Similar to the restricted case, here the tadpole-like orbits are more elongated in the opposite direction of the primary satellite. In this way, a greater number of stable systems are expected when we place the co-orbital satellites with $\theta > 60^{\circ}$ than otherwise. This feature explains the asymmetries in the distribution of stable conditions around $\theta = 60^{\circ}$ shown in Fig. \ref{fig:initial-setup}.

\begin{figure}
\begin{center}
\includegraphics[height=0.7\linewidth, width=\linewidth]{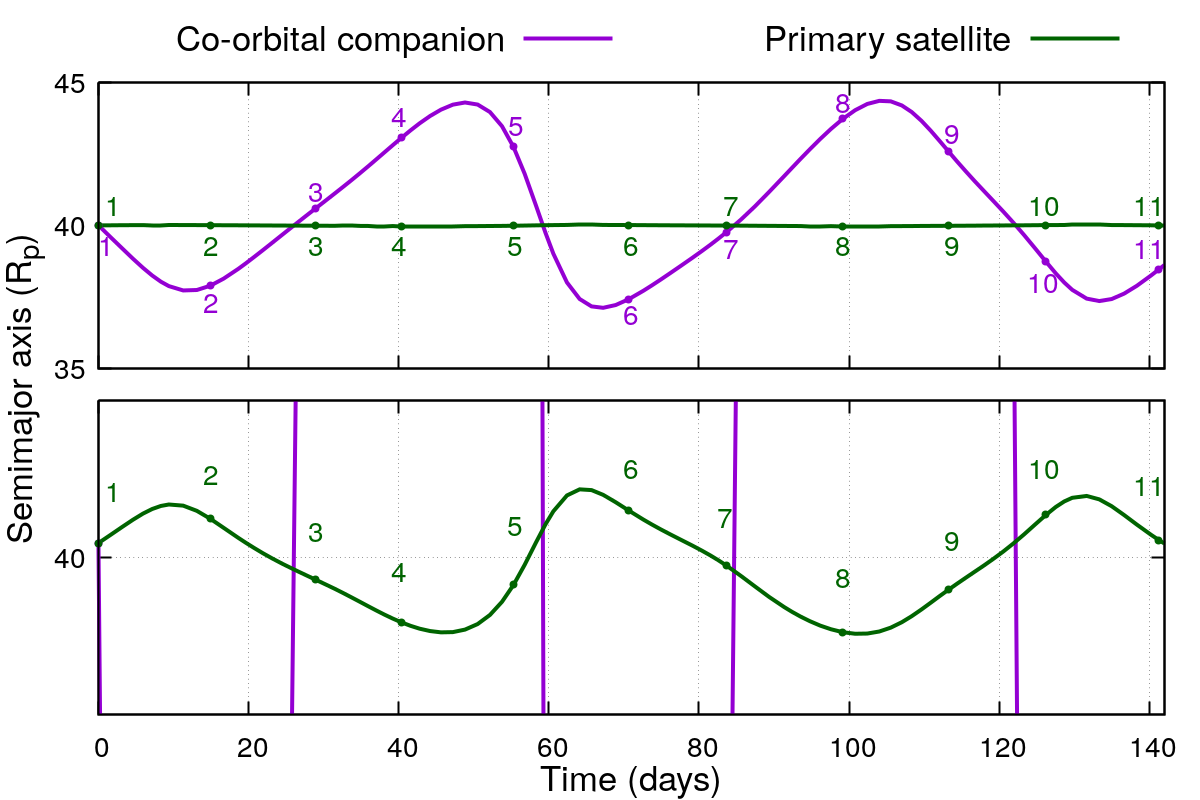}
\caption[Semimajor]{Evolution of the semimajor axis versus time of the satellites for the system Kepler-1625, where the co-orbital companion has $0.107$ $M_{\oplus}$ (Mars-size) and the satellites were initially $48^{\circ}$ apart from each other. The bottom panel is a zoom on the region with a semimajor axis between $39.95$ and $40.05$ $R_p$. The numbered points correspond to the same points presented in Fig. \ref{fig:motion}.}
\label{fig:zoom}
\end{center}
\end{figure}

The numbered points in both panels of Fig. \ref{fig:motion} represent the position of each satellite at the same time. If we follow these points, we notice that the satellites' motions are synchronized, such as both satellites crossed their initial semimajor axis at the same time (points of closest and farthest approach). Also, while one satellite is in the inner portion of its orbit (closer to the planet) the other is in the outer portion (farther from the planet) (Fig. \ref{fig:zoom}). At their closest approach, the satellites exchange angular momentum, such as the orbit of the smaller satellite shrinks while the orbit of the bigger satellite expands. The ballet performed by the satellites resembles the motion of Janus and Epimetheus in the Saturn system, but with both satellites having tadpole-like orbits (Epimetheus is in a horseshoe orbit).

The two motions that shaped the orbits of the satellites have an associated frequency since the epicentric and epicyclic motions have long and short periods, respectively. In the restricted three-body problem, the commensurability of these frequencies can give birth to resonances of libration, which may cause instabilities in the system \citep{Erdi-Sandor-2005}.
For the restricted three-body problem, we can define the mass parameter $\mu = M_1/(M_p+M_1)$, where $M_p$ and $M_1$ are the masses of the central planet and the orbiting massive body, respectively. The frequencies of motion of a particle around $L_4$ are given by \citet{Murray-1999},
\begin{align}\label{eq:freq1}
\lambda_{1,2} = \pm \dfrac{\sqrt{-1-\sqrt{1-27(1-\mu)\mu}}}{\sqrt{2}}
\end{align}
and
\begin{align}\label{eq:freq2}
\lambda_{3,4} = \pm \dfrac{\sqrt{-1+\sqrt{1-27(1-\mu)\mu}}}{\sqrt{2}},
\end{align}
where $\lambda_{1,2}$ is the frequency of the short-period epicyclic motion and $\lambda_{3,4}$ is the frequency of the long-period motion of the epicentre about $L_4$. 

Taking the ratio $\lambda_{1,2}/\lambda_{3,4}$, if we can find commensurabilities between the frequencies that can be written as the ratio of two integers, we will have resonances of librational frequencies. 

One should notice that these frequencies (Eqs. \ref{eq:freq1} and \ref{eq:freq2}) depend on the mass parameter of the system, such as different resonances will appear only for systems with specific values of $\mu$.

As shown by \citet{Erdi-Sandor-2005} (their Fig. 6), some librational frequencies, in their study the 2:1 and 3:1, can cause instabilities in co-orbital systems. They showed that around the 2:1 resonance, stable systems are not found for any value of eccentricity of the secondary body, thus creating an instability island for certain values of $\mu$. Also, the authors found that the 3:1 resonance causes the number of stable systems to decrease abruptly. However, they still found stability for cases with lower eccentricities of the secondary body. Other librational frequency resonances can be spotted in their work, for example, the 3:2, but similar to the 3:1 resonance, this commensurability only leads to a decrease in the number of stable systems, implying that stability in this case only can be found if the secondary body does not have an eccentric orbit ($e < 0.1$).

\begin{figure}
\begin{center}
\includegraphics[height=0.66\linewidth, width=1.0\linewidth]{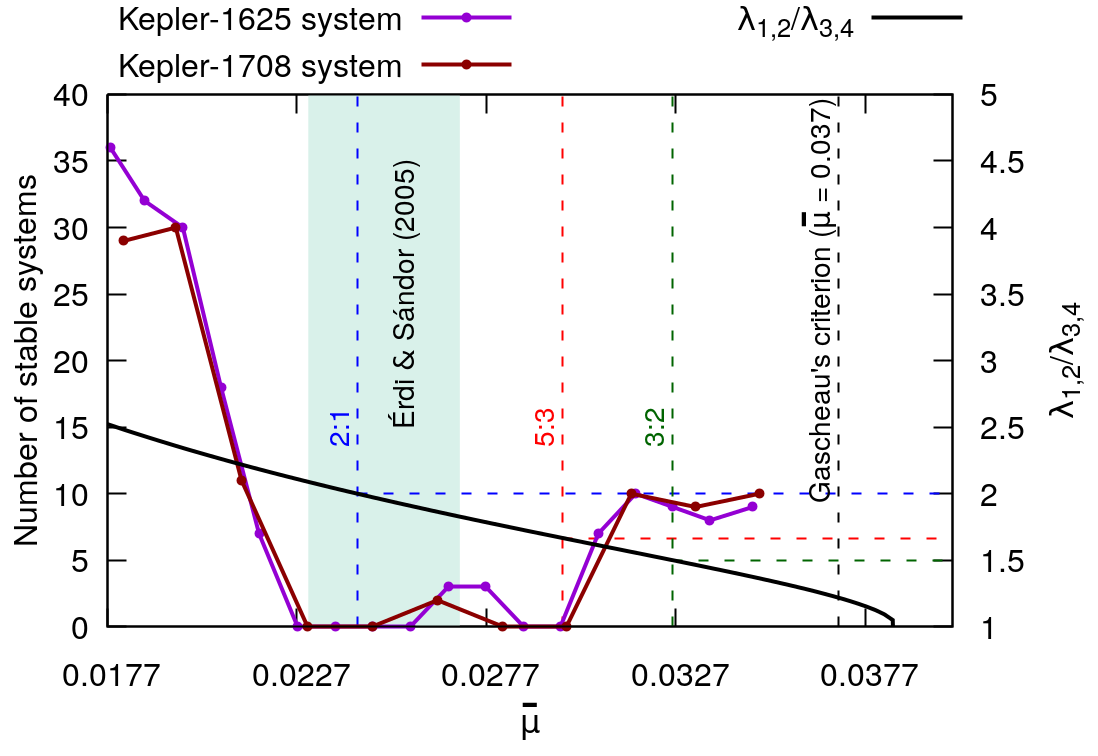}
\caption[Resonances]{Number of stable systems as a function of the mass parameter $\bar{\mu}$ for the system Kepler-1625 (purple dotted line) and Kepler-1708 (dark-red dotted line). The dots mark the value of $\bar{\mu}$ as we vary the mass of the co-orbital companion. The black curve is the ratio $\lambda_{1,2}/\lambda_{3,4}$ as a function of $\bar{\mu}$, the vertical black dashed line marks the Gascheau's criterion limit ($\bar{\mu} = 0.037$), the blue, red and green dashed lines denote the location of the 2:1, 5:3 and 3:2 librational frequency resonances for the restricted three-body problem and the region in cyan is an approximation of the region of instability found by \citet{Erdi-Sandor-2005} (their Fig. 6).}
\label{fig:gascheau}
\end{center}
\end{figure}

Some results from the restricted three-body problem are expected to hold in the non-restricted case. In Fig. \ref{fig:stability}, we have unstable regions for certain masses of the co-orbital companion. These results are similar to the ones found by \citet{Erdi-Sandor-2005}, thus the nature of the instability could be the same. 

To compare our results with the restricted case we define the mass parameter in our systems as
\begin{align}\label{eq:mass-parameter}
    \bar{\mu} = \dfrac{M_1+M_2}{M_p+M_1+M_2},
\end{align}
which is a generalization of the mass parameter considered in the restricted case. 

\citet{Gascheau-1843} and \citet{Routh-1874} found that, if the co-orbital bodies were in circular and coplanar motion, the Lagrangian points $L_4$ and $L_5$ will be linearly stable if 
\begin{align}\label{eq:gascheau}
    \dfrac{M_pM_1 + M_pM_2 + M_1M_2}{\left(M_p + M_1 + M_2\right)^2} < \dfrac{1}{27}.
\end{align}
Neglecting terms of second order and more in Eq. \ref{eq:gascheau}, we have \citep{Leleu-etal-2015}
\begin{align}\label{eq:leleu2015}
\dfrac{M_1 + M_2}{M_p + M_1 + M_2} \le \dfrac{1}{27}. 
\end{align}
One can see that the left-hand side of Eq. \ref{eq:leleu2015} is equal to the definition of $\bar{\mu}$ (Eq. \ref{eq:mass-parameter}). Thus, we have that $\bar{\mu} \le 1/27\sim 0.037$ (Gascheau's criterion). In our case, the motions of the two satellites are coplanar to the planet but only initially circular, i.e., the satellites can acquire non-negligible eccentricities during their evolution. In this way, Gascheau's criterion may not always apply, and the stability at $L_4$ is not guaranteed. In fact, \citet{Deprit-Deprit-Bartholome-1967} showed that for small values of eccentricity, the Gascheau's criterion applied to the restricted three-body problem (Eq. \ref{eq:gascheau} with $M_2 = 0$), $\mu < 0.0385$ (also known as Routh's critical mass ratio), makes the co-orbital region around the Lagrangian points unstable.

To estimate the locations of the librational frequency resonances in our systems, we calculate the ratio of the frequencies of motion, $\lambda_{1,2}/\lambda_{3,4}$ (Eqs. \ref{eq:freq1} and \ref{eq:freq2}), using the mass parameter $\bar{\mu}$ defined in Eq. \ref{eq:mass-parameter} for all our systems. We expect that this ratio will give us an approximation for the location of the resonances once the frequencies of motion are valid only for the restricted three-body problem.

In Fig. \ref{fig:gascheau} we show the number of stable systems as a function of the mass parameter $\bar{\mu}$ for the system Kepler-1625 (purple dotted line) and Kepler-1708 (dark-red dotted line), and the ratio $\lambda_{1,2}/\lambda_{3,4}$ as a function of $\bar{\mu}$ (represented by the black curve) which we used to find the libration frequency resonances 2:1 (blue dashed line), 5:3 (red dashed line) and 3:2 (green dashed line). To directly compare our results with \citet{Erdi-Sandor-2005}, we draw the region in cyan, representing an approximation of the instability region they found.   

From Fig. \ref{fig:gascheau} one can see that the island of instability found by \citet{Erdi-Sandor-2005} was recovered in our simulations. This instability is driven by the 2:1 libration frequency resonance. As we increase the mass of the co-orbital companions, and consequently the value of $\bar{\mu}$, we found stable satellites only near $L_4$. In these cases, the 2:1 resonance is still affecting the co-orbital satellites, but close to $L_4$, the perturbations are weaker.

Near the 5:3 resonance, we also found only unstable systems. This result agrees with the prediction of \citet{Erdi-Sandor-2005}, where the authors found only a few stable orbits for satellites in nearly circular motion. We also located the 3:2 libration frequency resonance. For this value of $\bar{\mu}$, equivalent to $M_2 = 16$ $M_{\oplus}$ (system Kepler-1625) and $M_2 \sim 4.5$ $M_{\oplus}$ (system Kepler-1708), we found eight and nine initial conditions that returned stable systems for the Kepler-1625 and the Kepler-1708 system, respectively. In all cases, both satellites sustained an almost circular orbit around the planet. These results are in good agreement with the findings of \citet{Erdi-Sandor-2005} as well. 

\subsection{Angular Instabilities}

In addition to the instabilities that appeared because of the variation in the mass of the co-orbital companion, we detected some smaller islands of instability inside larger islands of stability for certain angular separations. In the Kepler-1625 system, there are two cases of angular instabilities. These two cases appear when we set the co-orbital companion to: (i) a Mars-sized body initially with $\theta = 51^{\circ}$; and (ii) a 16-Earth masses body initially with $\theta = 56^{\circ}$. For case (i), the system becomes unstable after $\sim 1362$ years when a collision between the satellites takes place. In case (ii), the two satellites also collided, but after only $\sim 62$ years. For the Kepler-1708 system, when the co-orbital satellites have the same masses ($5$ $M_{\oplus})$ we found instability for two specific angles inside a stability island, $\theta = 54^{\circ}$ and $55^{\circ}$. In both cases, the satellites collided with each other before $2$ years.

Even though the nature of the above-mentioned instabilities is the same, to make the manuscript clearer we will separate the analysis of the Kepler-1625 and Kepler-1708 systems.

\subsubsection{Kepler-1625 system}
\begin{figure}
\begin{center}
\includegraphics[height=0.7\columnwidth, width=\columnwidth]{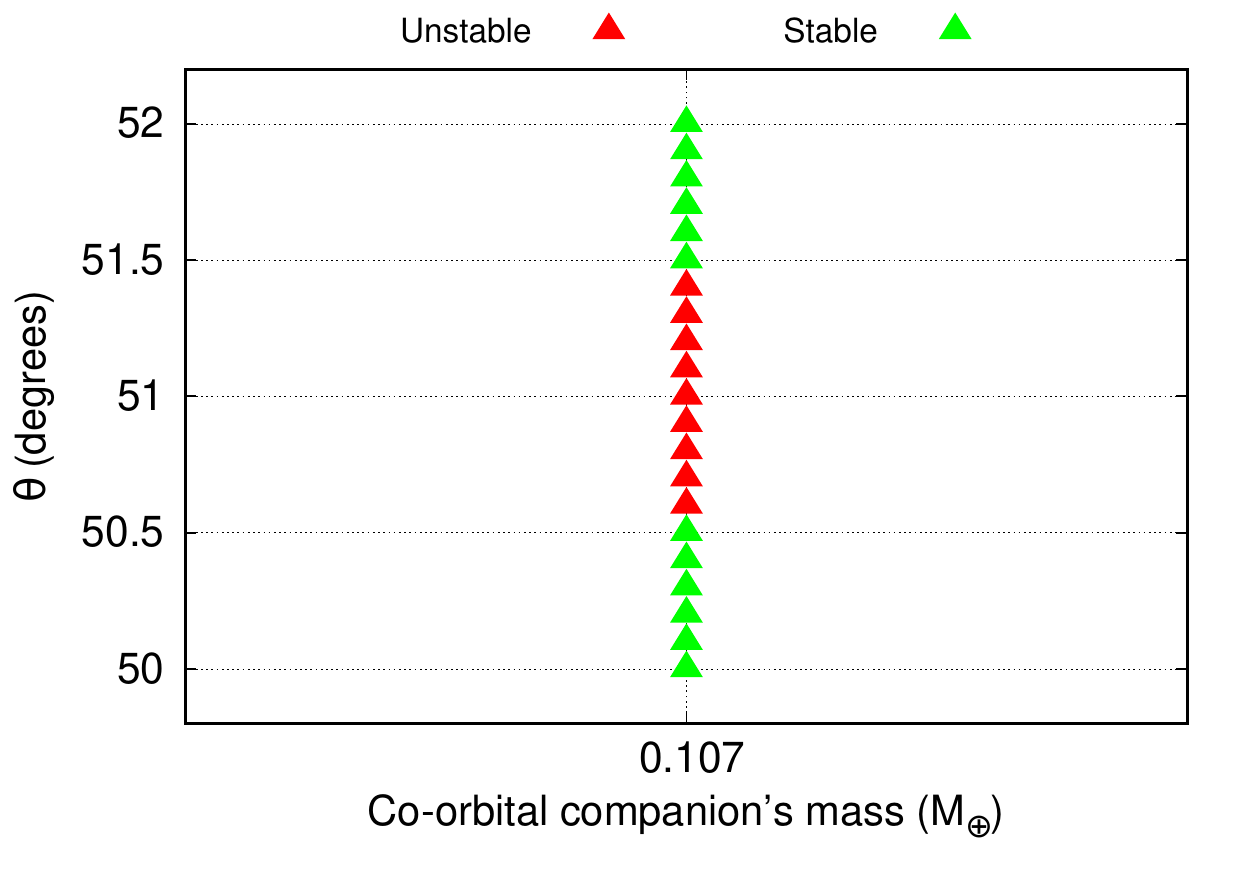}
\includegraphics[height=0.7\columnwidth, width=\columnwidth]{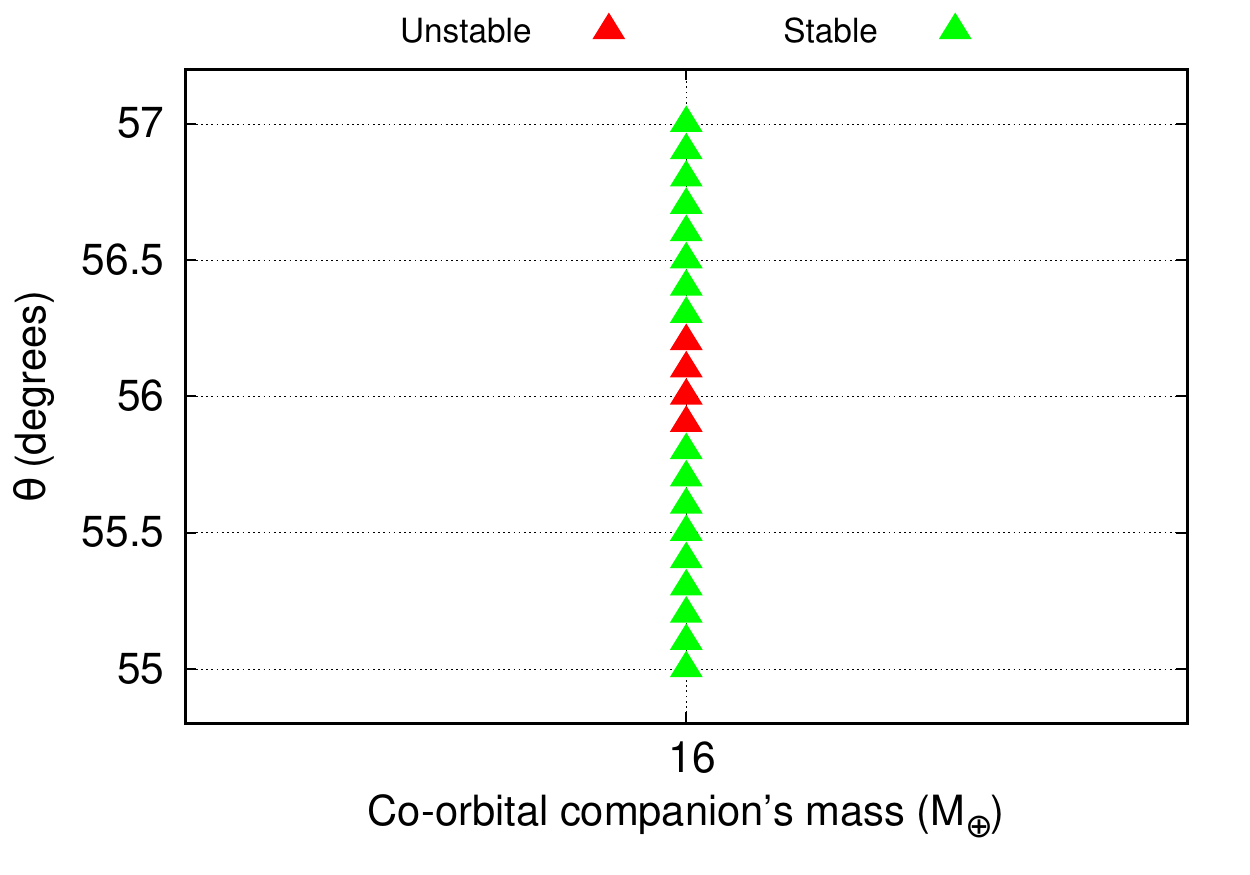}
\caption[Island of Instability]{Zoom on the angular separation of the isolated instabilities found in the Kepler-1625 system for a Mars-sized co-orbital companion initially at $\theta = 51^{\circ}$ (top panel) and for a 16-Earth masses co-orbital companion initially at $\theta = 56^{\circ}$ (bottom panel).}
\label{fig:island}
\end{center}
\end{figure}

Fig. \ref{fig:island} shows a zoom on the angular separation of the co-orbital satellites around the angles we found peculiar instabilities. For these cases, we performed more simulations using $\Delta \theta = 0.1^{\circ}$ to identify the extension of the instability islands. As one can see the instabilities are local, with small amplitudes (less than $1^{\circ}$). The structure of these instabilities suggests that initially, we had large stable islands, but due to resonant effects, these islands fragmented into smaller ones. In these cases, the unstable regions are dominated by chaotic motion (see Fig. 3 in \citet{Liberato-Winter-2020}).

One should notice that the librational frequency resonances we studied before cannot be responsible for these local instabilities since the frequencies given by Eqs. \ref{eq:freq1} and \ref{eq:freq2} are functions of the mass parameter of the system, while these new features are related to the angular separation of the satellites.

To find the potential resonances acting on $\theta$, we will study the time evolution of the angular separation between the co-orbital satellites. In this way, we can apply a Fast Fourier Transform (FFT) to isolate the dominant frequencies in the time series and investigate if resonances are causing the instabilities we observed.

\begin{figure}
\begin{center}
\includegraphics[height=0.7\columnwidth, width=\columnwidth]{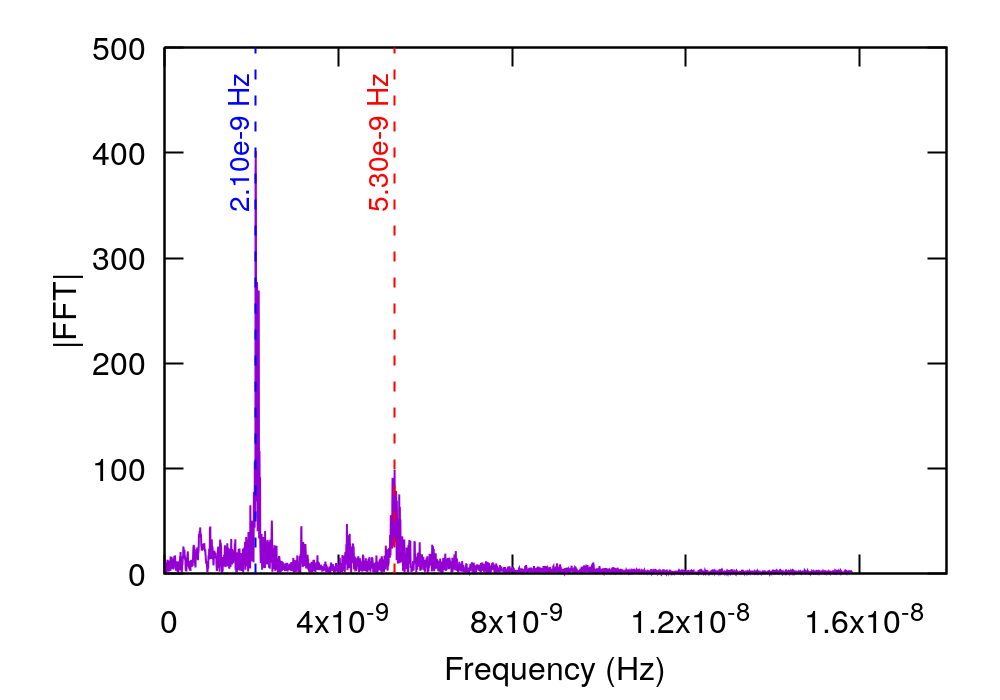}
\includegraphics[height=0.7\columnwidth, width=\columnwidth]{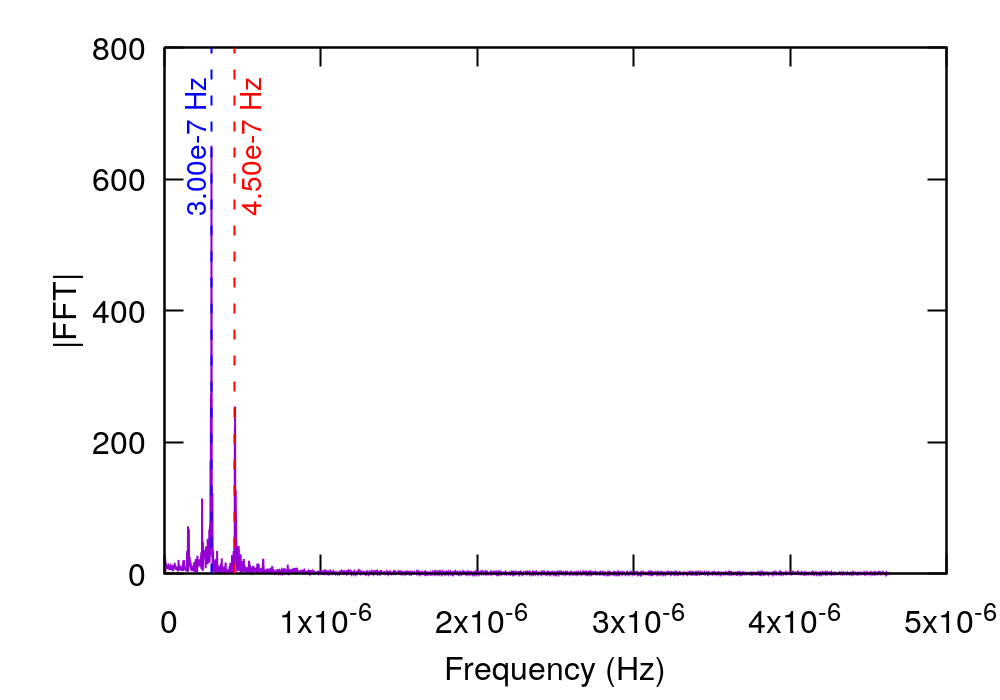}
\caption[Fast Fourier Transform]{Magnitude of the Fast Fourier Transform associated with the frequencies present in the evolution of the angular separation of the satellites in the Kepler-1625 system. Top panel: The co-orbital satellite is a Mars-sized body with an initial angular separation of $51^{\circ}$. Bottom panel: The co-orbital satellite is a 16-Earth masses body with an initial angular separation of $56^{\circ}$.}
\label{fig:fft}
\end{center}
\end{figure}

Fig. \ref{fig:fft} shows the magnitude of the FFT associated with the frequencies present in the evolution of the angular separation between the co-orbital satellites. 

On the top panel of Fig. \ref{fig:fft}, we have the FFT analysis of the frequencies of $\theta$ for the co-orbital pair formed by the primary satellite and the Mars-sized companion. As one can see, there are two peaks of magnitude around $2.10\times10^{-9}$ $Hz$ and $5.30\times10^{-9}$ $Hz$, representing the two dominant frequencies in $\theta$. Taking the ratio of these two frequencies, we find approximately a $5/2$ commensurability, which can be understood as a 5:2 resonance between the libration of the co-orbital satellite about $L_4$ and the angular motion period of the satellites. This third-order resonance was responsible for increasing the eccentricity of the co-orbital companion. In this way, the orbits of the satellites crossed and led to a collision between the bodies. 

The same discussion applies to the co-orbital pair of the primary satellite and the 16-Earth masses companion with initial angular separation $\theta = 56^{\circ}$. The FFT analysis revealed the frequencies $3.00\times10^{-7}$ $Hz$ and $4.50\times10^{-7}$ $Hz$ as the dominant ones on the evolution of $\theta$. Once again, comparing these frequencies, we find a 3:2 commensurability. This particular resonance suddenly increased the eccentricity of both satellites resulting in a collision.
\subsubsection{Kepler-1708 system}

For the Kepler-1708 system, we found an island of instability inside a greater island of stable initial conditions only for the systems where the satellites have the same masses, $5$ $M_{\oplus}$. In this case, the satellites are stable when their initial angular separation is $\theta = 53^{\circ}$ and for $\theta = 56^{\circ} - 64^{\circ}$, leaving a gap of unstable conditions for $\theta = 54^{\circ}$ and $55^{\circ}$.

\begin{figure}
\begin{center}
\includegraphics[height=0.7\columnwidth, width=\columnwidth]{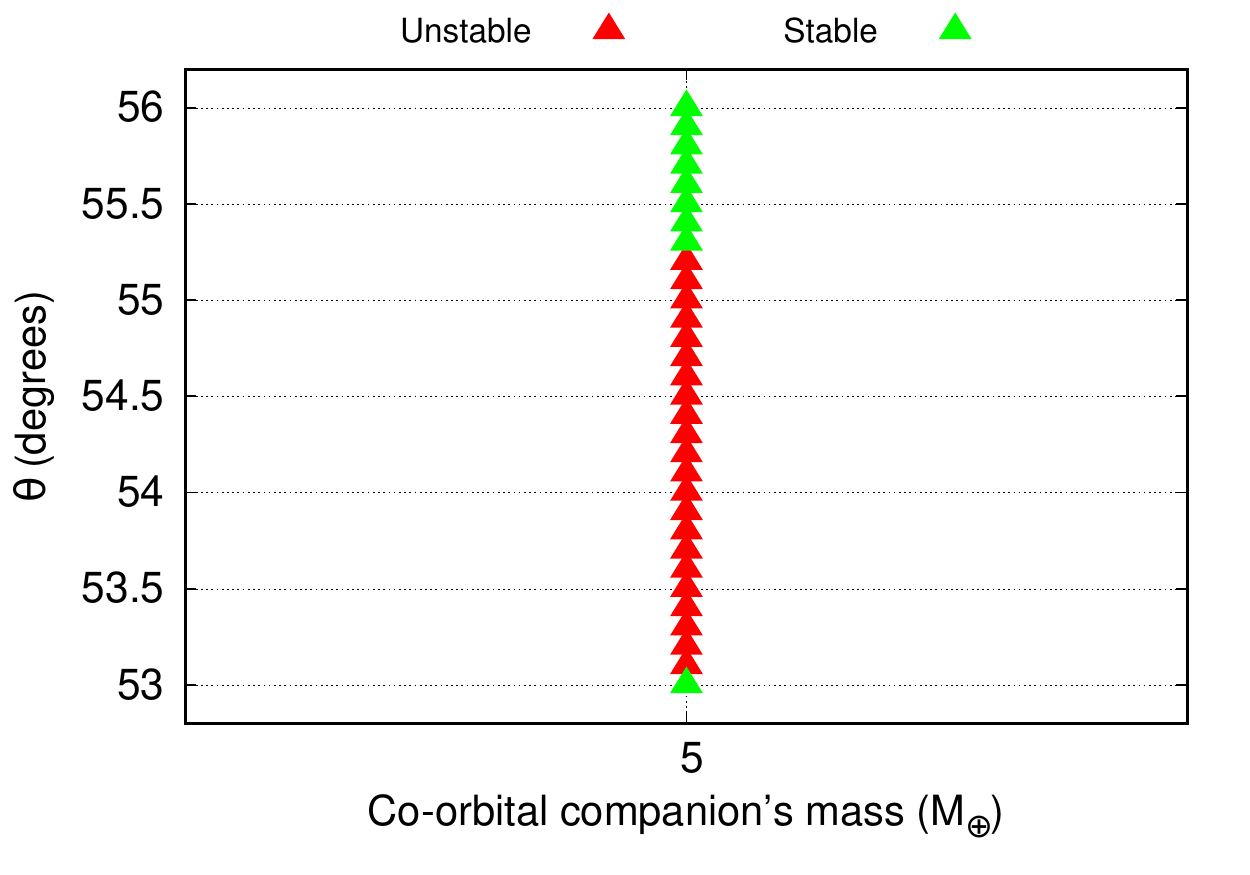}
\caption[Island of Instability Kepler-1708]{Zoom on the angular separation of the isolated instabilities found in the Kepler-1708 system for a 5-Earth masses co-orbital companion initially located at  $\theta = 54^{\circ}$ and $55^{\circ}$.}
\label{fig:zoom-Kepler1708}
\end{center}
\end{figure}

\begin{figure}
\begin{center}
\includegraphics[height=0.7\columnwidth, width=\columnwidth]{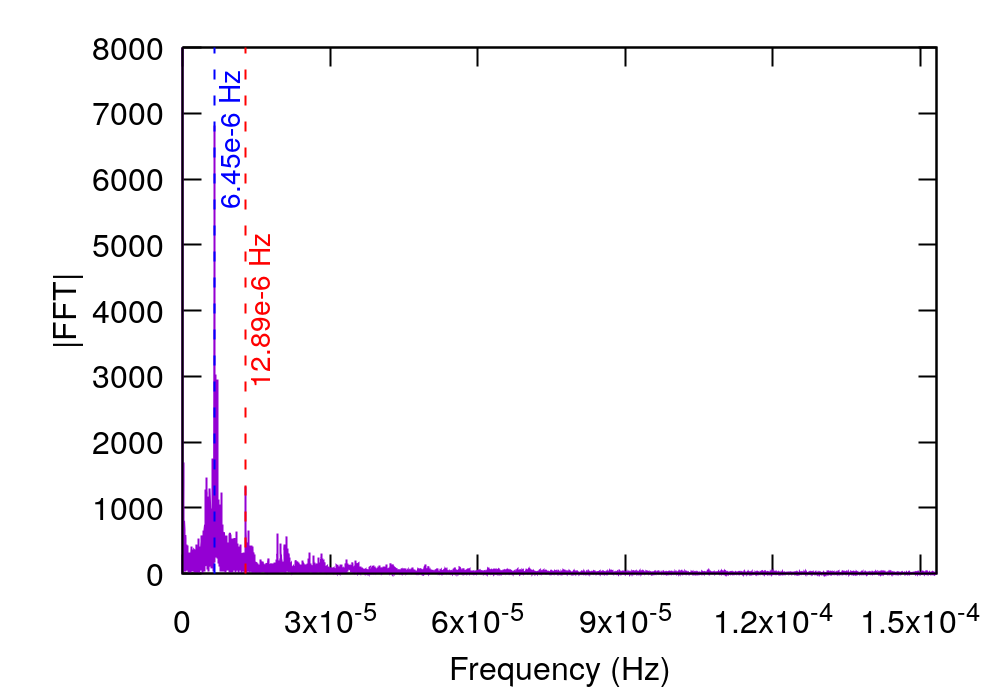}
\includegraphics[height=0.7\columnwidth, width=\columnwidth]{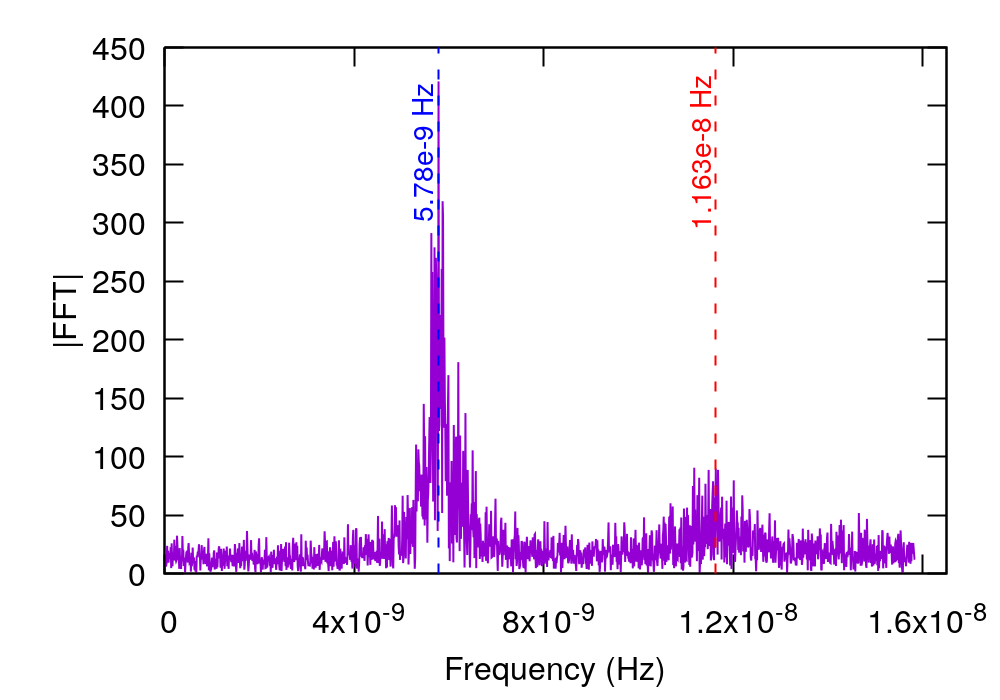}
\caption[Fast Fourier Transform Kepler-1708]{Magnitude of the Fast Fourier Transform associated with the frequencies present in the evolution of the angular separation of the satellites in the Kepler-1708 system. Top panel: Initial angular separation of $54^{\circ}$. Bottom panel: Initial angular separation of $56^{\circ}$. In both cases the satellites have the same masses, $5$ $M_{\oplus}$}
\label{fig:fft-54-55}
\end{center}
\end{figure}

To precisely locate the border of this instability, we follow the approach used for the Kepler-1625 system, first refining the initial conditions between $\theta = 53^{\circ}$ and $55^{\circ}$ using $\Delta \theta = 0.1^{\circ}$ and then applying an FFT analysis for the frequencies of $\theta$. Differently from the previous cases, for the Kepler-1708 system, the amplitude of the angular instability is wider than $1^{\circ}$ (Fig. \ref{fig:zoom-Kepler1708}) extending from $53.1^{\circ}$ to $55.2^{\circ}$. In most cases, the satellites collided with each other within one year, which could point toward the presence of strong resonances acting over the satellites' angular separation.

Fig. \ref{fig:fft-54-55} shows the FFT analysis of the frequencies in the evolution of $\theta$ for the initial separations of $54^{\circ}$ (top panel) and $55^{\circ}$ (bottom panel), respectively. In both cases, taking the ratio between the dominant frequencies, we find approximately 2, which indicates the proximity of the 2:1 resonance. This first-order resonance abruptly increases the eccentricity of the satellite leading to an almost instantaneous collision.
\begin{figure*}
\begin{center}
\includegraphics[height=0.32\linewidth, width=0.49\linewidth]{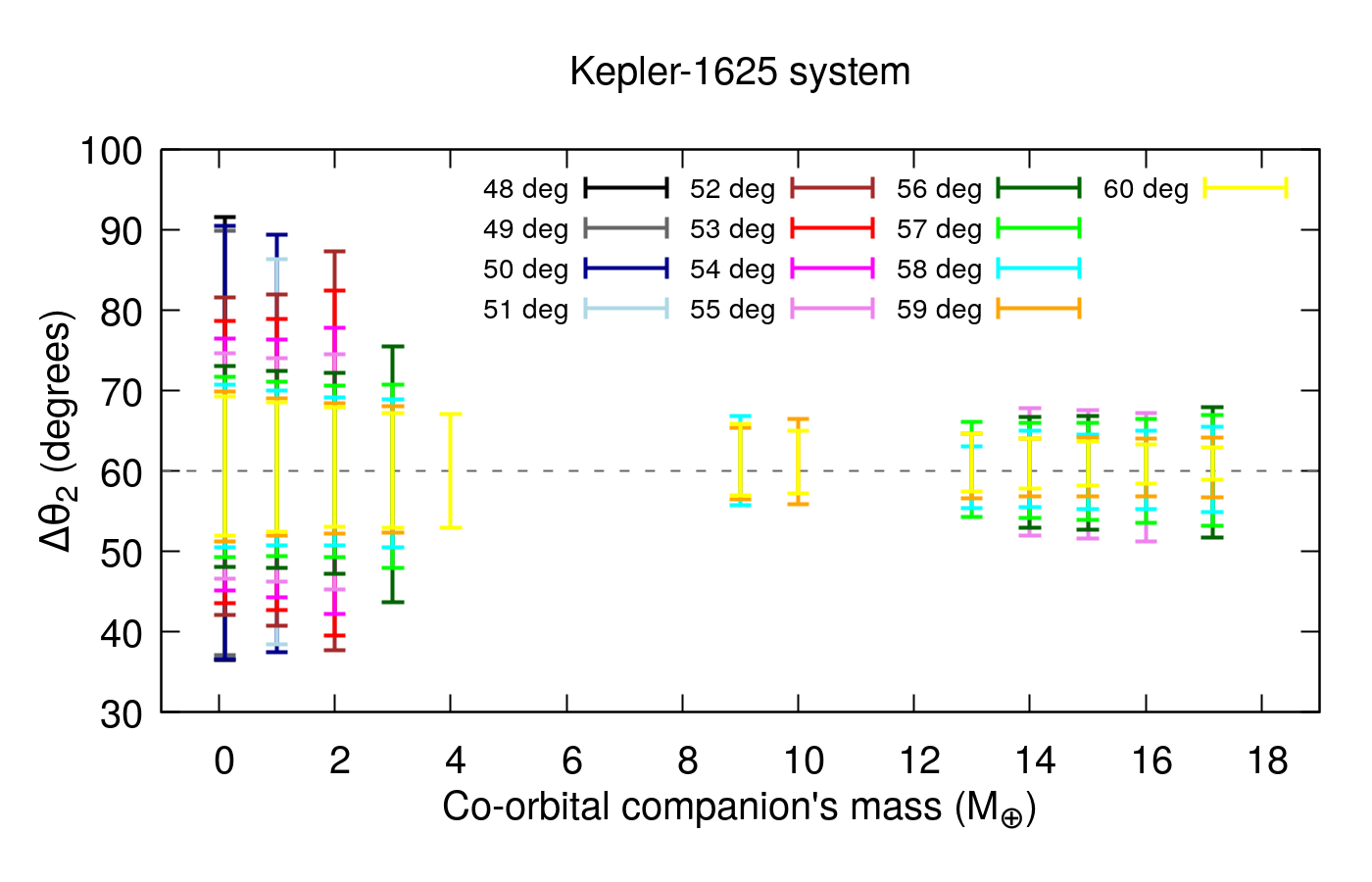}
\includegraphics[height=0.32\linewidth, width=0.49\linewidth]{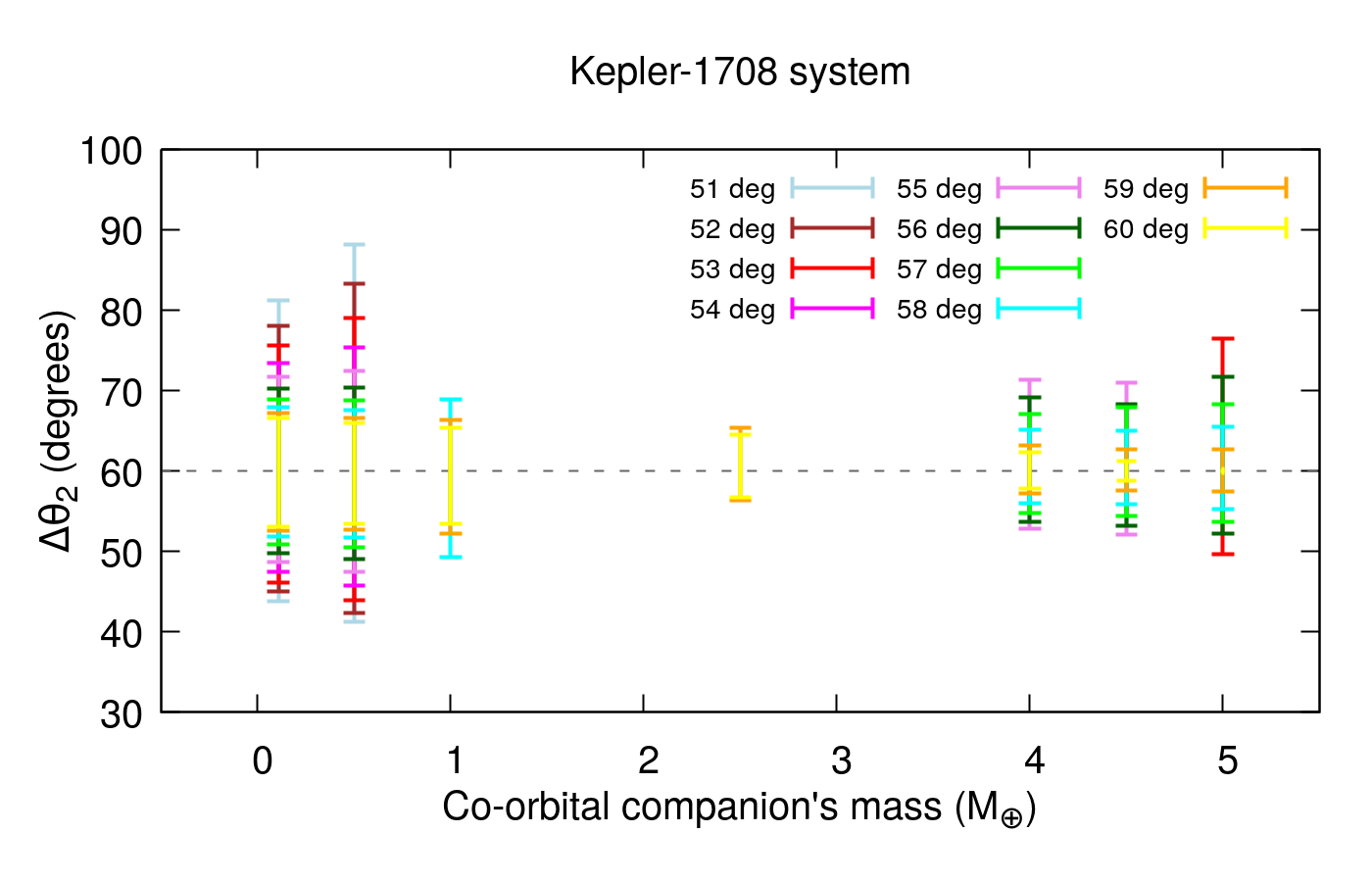}
\caption[Amplitude]{Amplitude of libration of the co-orbital companions ($\Delta \theta_2$) vs their masses for the systems Kepler-1625 (top panel) and Kepler-1708 (bottom panel). The colour scheme represents the initial $\theta$ of each satellite.}
\label{fig:amp}
\end{center}
\end{figure*}

\subsection{Amplitude of Motion}

As shown before, the orbits of the primary satellite and its co-orbital companion have a tadpole-like shape. However, we did not address the amplitude of these motions since we only showed one example (a system with a Mars-like co-orbital companion with an initial angular separation of $\theta = 48^{\circ}$ in the Kepler-1625 system).

The amplitude of the satellites' orbits depends on the magnitude of the perturbations they will receive from the other satellite and the planet. On the other hand, these perturbations are proportional to the mass and initial angular separation of the co-orbital pair. For example, less massive co-orbital companions will be more affected by perturbations from the primary satellite than more massive companions. Also, as the $L_4$ is an equilibrium point, such as the net force at this point is minimum, at this location, the amplitude of the satellites' motion is expected to be small. 

In Fig. \ref{fig:amp} we show the amplitude of motion of the co-orbital companion satellites ($\Delta \theta_2$) for each initial angular separation for the Kepler-1625 (left panel) and Kepler-1708 (right panel) system, respectively. For better visualization, we only present the cases for $\theta\le60^{\circ}$, but for greater separations, the behaviour follows the same pattern. The primary satellite always starts at $\theta = 0^{\circ}$. Thus, its angular displacement is not significant compared to the motion of the co-orbital companion. In this way, we will neglect the analysis of this motion. 

As expected, the motions of the co-orbital companions are around $L_4$ ($\theta = 60^{\circ}$), although not necessarily symmetrical. Also, as the mass parameter $\bar{\mu}$ decreases, stable orbits with larger amplitudes of libration are possible \citep{Leleu-etal-2018}. In our simulations, we did not find satellites in horseshoe orbits. \citet{Roberts-2002} showed throughout analytic calculations that horseshoe configurations in the general three-body problem are stable only for $\bar{\mu}\le3\times10^{-4}$, which is not the case for our systems. For the restricted three-body problem, numerical simulations found this limit to be $\mu \le 9.5\times10^{-4}$
\citep{Liberato-Winter-2020}.

Stable co-orbital companions initially farther from $L_4$ presented a larger amplitude of motion because of the perturbations they received from the primary satellite at the moment of their closest approach. If the satellites are too close to each other, this close encounter may result in collisions or ejections of the less massive body. On the other hand, for satellites with initial $\theta>60^{\circ}$, the amplitude of libration is similar to the ones presented here.


\section{Results for the complete system: Star-Planet-Satellite-Co-orbital companion}\label{sfour}

In this section, we study the stability of co-orbital exomoons taking into account the star of the systems. The initial conditions for the planet-satellites systems will be the same as in Sec. \ref{sthree}, only adding the star as the central body and considering the respective planet with the semimajor axis given in Tab. \ref{tab:properties}. The planets will be considered in circular and coplanar orbits. For the system Kepler-1625, the planet is predicted to have a non-neglectable inclination \citet{Teachey-etal-2018}, in this case, we will also simulate the case with $I_p = 45^{\circ}$.

In adding the star, we will study the influence of the satellites on the planet's motion. In this way, we can find the planet's TTVs characterized by the presence of co-orbital exomoons.

In the following, we analyze the results for the Kepler-1625 and Kepler-1708 systems separately.

\subsection{Kepler-1625 system}

For the Kepler-1625 system considering the star, all the simulated cases ended with unstable systems regardless of the mass of the co-orbital companion, its initial angular position, or the inclination of the planet. In Tab. \ref{tab:results-Kepler1625}, we present a summary of the outcomes of the simulations separated by the co-orbital companions' masses.

Even though the satellites are well inside the Hill radius, $a_s \sim 0.264$ $R_{H,p}$, the gravitational perturbations of the star are strong enough to disrupt the initial co-orbital architecture of the satellites. The systems fall apart usually within a few centuries, resulting in collisions between the planet and the satellites, collisions between the satellites, ejections of one of the moons, or the exomoon leaving its planetocentric orbit and assuming an orbit around the star, thus becoming a planet or a \textit{ploonet} \citep{Sucerquia-etal-2019}.

\begin{table*}
\begin{center}
\caption[Results system Kepler-1625]{Results of our simulations for the system Kepler-1625 considering the star. For each column, we have the systems separated by the mass of the co-orbital companion ($M_2$), and the number and percentage of systems with collisions between the planet and the co-orbital companion (Coll. $M_p - M_2$), collisions between the satellites (Coll. $M_1 - M_2$), collisions between the planet and Kepler-1625 b-I (Coll. $M_p - M_2$), ejections of the co-orbital companion ($M_2$ Ejected), ejections of Kepler-1625 b-I ($M_1$ Ejected), and ploonets. In the last row, we have the summation of all simulated cases and their respective percentages.}
\begin{tabular}{ccccccc}\hline 
 $M_2$          & Coll. $M_p - M_2$ & Coll. $M_1 - M_2$ & Coll. $M_p - M_1$ & $M_2$ Ejected & $M_1$ Ejected & Ploonet \\
 $M_{\oplus}$   &                  &                   &                   &               &               & \\ \hline	
$0.107$ & $ 26\;(\sim 42.6\%) $ & $ 28\;(\sim 45.9\%) $ & $ 7\;(\sim 11.5\%) $ & $ - $ & $ - $ & $ - $  \\ \hline
$1.0$ & $ 24\;(\sim 39.3\%) $ & $ 31\;(\sim 50.8\%) $ & $ 6\;(\sim 9.8\%) $ & $ - $ & $ - $ & $ - $  \\ \hline
$2.0$ & $ 24\;(\sim 39.3\%) $ & $ 33\;(\sim 54.1\%) $ & $ 4\;(\sim 6.6\%) $ & $ - $ & $ - $ & $ - $  \\ \hline
$3.0$ & $ 22\;(\sim 36.1\%) $ & $ 33\;(\sim 54.1\%) $ & $ 5\;(\sim 8.2\%) $ & $ - $ & $ - $ & $ 1\;(\sim 1.6\%)$  \\ \hline
$4.0$ & $ 20\;(\sim 32.8\%) $ & $ 32\;(\sim 52.5\%) $ & $ 9\;(\sim 14.8\%) $ & $ - $ & $ - $ & $ - $  \\ \hline
$5.0$ & $ 24\;(\sim 39.3\%) $ & $ 29\;(\sim 47.5\%) $ & $ 7\;(\sim 11.5\%) $ & $ 1\;(\sim 1.6\%) $ & $ - $ & $ - $  \\ \hline
$6.0$ & $ 23\;(\sim 37.7\%) $ & $ 28\;(\sim 45.9\%) $ & $ 8\;(\sim 13.1\%) $ & $ 2\;(\sim 3.3\%)$ & $ - $ & $ - $  \\ \hline
$7.0$ & $ 16\;(\sim 26.2\%) $ & $ 35\;(\sim 57.4\%) $ & $ 7\;(\sim 11.5\%) $ & $ 3\;(\sim 4.9\%) $ & $ - $ & $ - $  \\ \hline
$8.0$ & $ 16\;(\sim 26.2\%) $ & $ 32\;(\sim 52.4\%) $ & $ 3\;(\sim 4.9\%) $ & $ 9\;(\sim 14.8\%) $ & $ 1\;(\sim 1.6\%) $ & $ - $  \\ \hline
$9.0$ & $ 16\;(\sim 26.2\%) $ & $ 36\;(\sim 59.0\%) $ & $ 4\;(\sim 6.6\%) $ & $ 4\;(\sim 6.6\%) $ & $ 1\;(\sim 1.6\%) $ & $ - $  \\ \hline
$10.0$ & $ 10\;(\sim 16.4\%) $ & $ 42\;(\sim 68.9\%) $ & $ 2\;(\sim 3.3\%) $ & $ 6\;(\sim 9.8\%) $ & $ 1\;(\sim 1.6\%) $ & $ - $  \\ \hline
$11.0$ & $ 14\;(\sim 22.9\%) $ & $ 36\;(\sim 59.0\%) $ & $ 5\;(\sim 8.2\%) $ & $ 5\;(\sim 8.2\%) $ & $ 1\;(\sim 1.6\%) $ & $ - $  \\ \hline
$12.0$ & $ 12\;(\sim 19.7\%) $ & $ 43\;(\sim 70.5\%) $ & $ 2\;(\sim 3.3\%) $ & $ 1\;(\sim 1.6\%) $ & $ 1\;(\sim 4.9\%) $ & $ - $  \\ \hline
$13.0$ & $ 10\;(\sim 16.4\%) $ & $ 39\;(\sim 63.9\%) $ & $ - $ & $ 7\;(\sim 11.5\%) $ & $ 5\;(\sim 8.2\%) $ & $ - $  \\ \hline
$14.0$ & $ 13\;(\sim 21.3\%) $ & $ 36\;(\sim 59.0\%) $ & $ 2\;(\sim 3.3\%) $ & $ 7\;(\sim 11.5\%) $ & $ 3\;(\sim 4.9\%) $ & $ - $  \\ \hline
$15.0$ & $ 17\;(\sim 27.9\%) $ & $ 29\;(\sim 47.5\%) $ & $ 1\;(\sim 1.6\%) $ & $ 8\;(\sim 13.1\%) $ & $ 6\;(\sim 9.8\%) $ & $ - $  \\ \hline
$16.0$ & $ 10\;(\sim 16.4\%) $ & $ 37\;(\sim 60.7\%) $ & $ 2\;(\sim 3.3\%) $ & $ 9\;(\sim 14.8\%) $ & $ 3\;(\sim 4.9\%) $ & $ - $  \\ \hline
$17.15$ & $ 3\;(\sim 4.9\%) $ & $ 45\;(\sim 73.8\%) $ & $ 1\;(\sim 1.6\%) $ & $ 10\;(\sim 16.4\%) $ & $ 2\;(\sim 3.3\%) $ & $ - $  \\ \hline
\textbf{Total}  & $\mathbf{300\;(\sim 27.3\%)} $ & $ \mathbf{624\;(\sim 56.8\%)} $ & $ \mathbf{75\;(\sim 6.8\%)} $ & $ \mathbf{72\;(\sim 6.6\%)} $ & $ \mathbf{26\;(\sim 2.4\%)} $ & $ \mathbf{1\;(\sim 0.09\%)}$  \\ \hline
\end{tabular}
\end{center}
\label{tab:results-Kepler1625}
\end{table*}

From Tab. \ref{tab:results-Kepler1625}, one can see that the most common outcome of our simulations was the collision between the satellites, especially for more massive co-orbital companions. This result is expected in the case of unstable systems since the satellites share the same orbit. Also, we point out the high number of systems ending with the collision between the co-orbital companion and the planet. In this case, the satellite is stripped from its co-orbital orbit, suffers an eccentricity excitation, and collides with the parent body.

To consider the possibility of satellites being ejected from the system, we set the distance of $300$ au from the star as the maximum distance a body could reach. Former satellites whose orbits extend beyond this limit are considered ejected from the system. Satellites were ejected only for systems with $5$-Earth masses satellites as co-orbital companions. For these systems, the close encounters between satellites are very energetic, resulting in the ejection of one of the satellites, usually the smaller one. Although, we also found cases in which the most massive satellite was the first body ejected from the system.

For the system formed with a $3$-Earth masses co-orbital companion initially at $\theta = 65^{\circ}$, we found that the secondary satellite was ejected from its planetocentric orbit in an inner heliocentric orbit. However, the satellite did not collide with the star and survived as a detached moon, or \textit{ploonet} as named by \citep{Sucerquia-etal-2019}. We follow the evolution of this body for $45$ thousand years, and the \textit{ploonet} remained in a stable configuration with the rest of the system. It is out of the scope of this work to investigate in detail the long-term evolution of this body, especially because it represented only $~0.09\%$ of the total sample. However, as shown in \citet{Hansen-2022}, unbound moons detached from their planet through tidally driven outward migration are likely to collide with the parental body within millions of years. In this way, the \textit{ploonet} we found could have the same fate long-term.

\subsection{Kepler-1708 system}

\subsubsection{Stability}

Different from the Kepler-1625 system, adding the star of the system Kepler-1708 did not significantly change our results regarding the stability of the co-orbital satellites.

In Tab. \ref{tab:results-Kepler1708}, we present a summary of the stable system for the cases with and without the star. As one can see, small differences between the results appear only on the border of the stability regions for some systems, while most of the results from Sec. \ref{sthree} are recovered. In this case, we argue that the gravitational influences of the star are minor or even negligible for the Kepler-1708 system.

\begin{table}
\begin{center}
\caption[Results system Kepler-1708]{Stable initial conditions for the system Kepler-1708 for the cases with and without the star. The results are presented as a function of the mass of the co-orbital companion ($M_2$).}
\begin{tabular}{ccc}\hline 
 $M_2$ & $\theta$ without the star    & $\theta$ with the star \\
 $M_{\oplus}$    & degrees   & degrees\\ \hline	
$0.107$ & $ 51-79 $ & $ 50-79 $ \\ \hline
$0.5$ & $ 51-80 $ & $ 51-80 $ \\ \hline
$1.0$ & $ 58-68 $ & $ 59-68 $ \\ \hline
$1.5$ & $ - $ & $ - $ \\ \hline
$2.0$ & $ - $ & $ - $ \\ \hline
$2.5$ & $ 59-60 $ & $ 59-60 $ \\ \hline
$3.0$ & $ - $ & $ - $ \\ \hline
$3.5$ & $ - $ & $ - $ \\ \hline
$4.0$ & $ 55-64 $ & $ 56-64 $ \\ \hline
$4.5$ & $ 55-63 $ & $ 55-63 $ \\ \hline
$5.0$ & $ 53, 56-64 $ & $ 53, 56-64 $ \\ \hline
\end{tabular}
\end{center}
\label{tab:results-Kepler1708}
\end{table}

There are mainly two reasons for the star not being relevant for the stability of co-orbital satellites on the Kepler-1708 system: (i) the star-planet separation; (ii) the position of the satellites related to the planetary Hill radius.

Kepler-1708 b is a cool-giant with a semimajor axis of $a_p = 1.64$ au. As the gravitational force of a body into another is inversely proportional to the distance between them, the influence of the star over the orbit of the planet, and consequently over its satellites, is less significant than what is expected for a close-in planet. For example, in the Kepler-1625 system, the star-planet distance is $0.87$ au and we found that co-orbital satellites are unstable if the star is considered.

The exomoon candidate Kepler-1708 b-I is predicted to have a semimajor axis of $a_s \sim 0.047$ $R_{H,p}$. At this distance, the gravitational force of the star is supplanted by the presence of the planet. Thus, the satellites will not feel the presence of the star. For the Kepler-1625 system, the satellites are farther from the planet ($~0.264$ $R_{H,p}$), while the planet is closer to the star. The combination of these two features jeopardized the presence of massive co-orbital satellites to Kepler-1625 b-I.

\subsubsection{Transit Timing Variations considering co-orbital exomoons}

Several authors proposed that TTVs could be used to indirectly detect the presence of more bodies in an exoplanetary system, planets \citep{Holman-Murray-2005, Agol-etal-2005, Nesvorny-Vokrouhlicky-2014, Agol-Fabrycky-2018} or exomoons \citep{Sartoretti-Schneider-1999, Szabo-etal-2006, Simon-etal-2007, Kipping-2021}. This indirect effect manifests itself as fluctuations in the timing of planetary transits and can be used to infer the presence of planets and moons in systems where at least one planet is known to be transiting.

In this section, we present a TTV analysis for the stable systems with co-orbital exomoons for the Kepler-1708 system.

Our synthetic TTVs will be constructed as follows: (i) we set-up a coplanar system, $(x,y)$, with the star in the origin of the system. The planet is positioned at $(a_p,0)$, and the satellites are placed around the planet; (ii) we define that the observer is located along the positive direction of the $x$-axis; (iii) the system will be integrated forward in time, with the planet motion being restricted to the $(x,y)$ plane and anticlockwise; (iv) at each half day interval we will verify if the planet crossed the $x$-axis from negative $y$ to positive $y$. If this is the case, we take the time before and after the passage through $x$ and apply a bisection method to precisely find the time of crossing, which we define as a transit time; (v) we stop the simulation when $200$ transits are obtained; (vi) at last, we remove the linear trend from our transit times applying a linear least square fit to our data. This process is done for systems considering only Kepler-1708 b-I and systems with the exomoon candidate and its co-orbital companion, such as we can measure the contribution of the co-orbital companion to the planet's TTV. The simulations are performed using the IAS15 integrator from the REBOUND \footnote{We opted to use REBOUND instead of POSIDONIUS here because REBOUND allows us to control the integration time and timestep more easily than POSIDONIUS.}. 

\begin{figure}
\begin{center}
\includegraphics[height=0.75\linewidth, width=\linewidth]{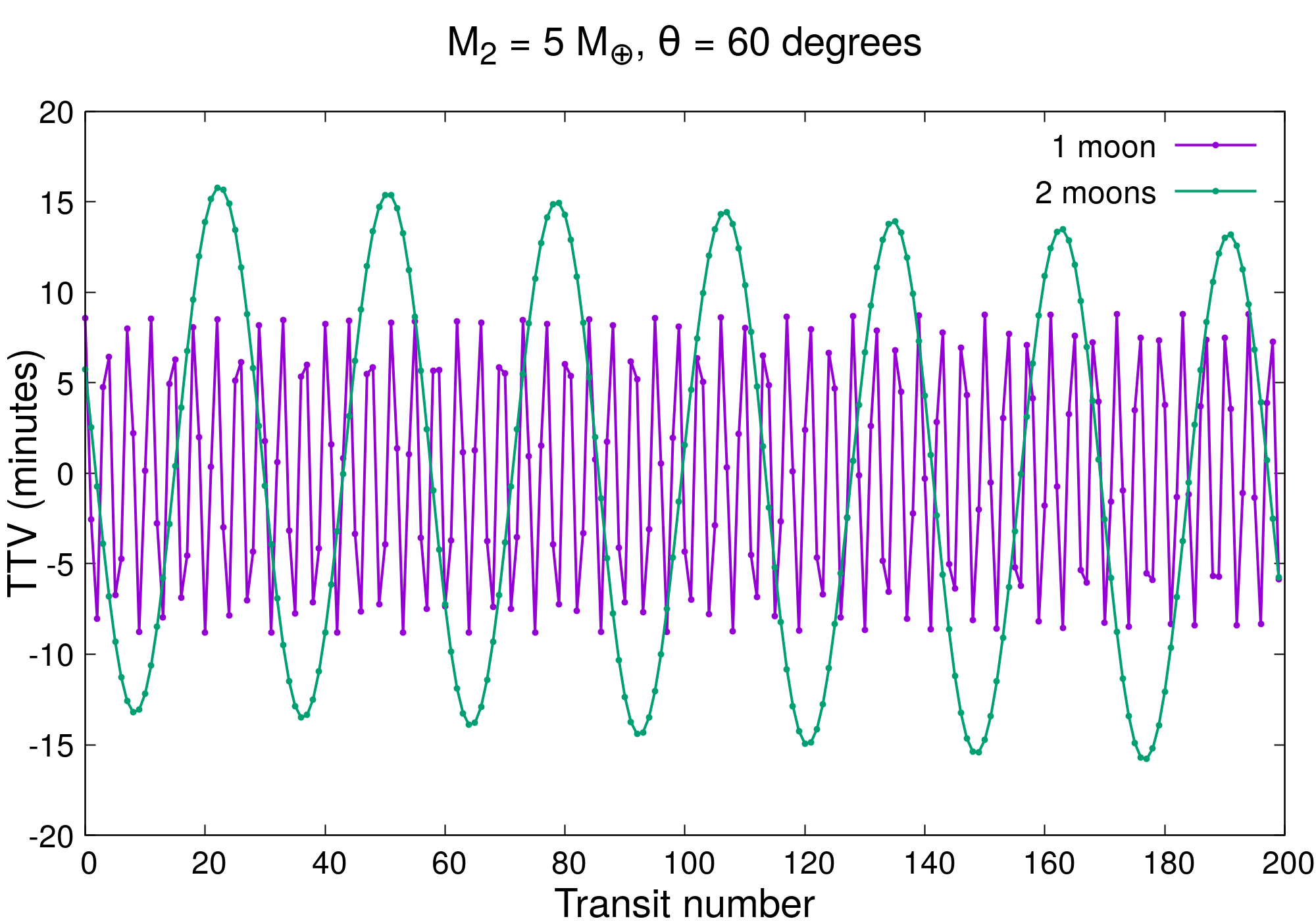}
\caption[TTV]{Transit timing variations for a planet with only one moon (purple) and for a planet with a co-orbital pair of satellites (green). The satellites have the same masses, $5$ $M_{\oplus}$, and they are initially $60^{\circ}$ apart from each other.} 
\label{fig:TTV-5-Earth}
\end{center}
\end{figure}

In Fig. \ref{fig:TTV-5-Earth}, we compare the TTVs generated by a system with only one moon and a system with a $5$ $M_{\oplus}$ co-orbital companion. As one can see, the amplitude of the TTV increased by more than 5 minutes with the addition of a co-orbital satellite with the same mass as the primary satellite. This variation is significant since the original amplitude of the TTV, considering only one moon, is smaller than 9 minutes, which is an expected outcome. As shown by \citet{Kipping-2009a}, the amplitude of the TTV is proportional to the mass of the exomoons. Also, we found that the presence of a co-orbital moon increased the periodicity of the TTVs, allowing us to draw a smoother fit for our data. 

For less massive co-orbital companions, the effects on the planet's TTV are minor. For example, for a Mars-like companion, the amplitude of the TTV only increased about 0.1 minutes.

We also investigated the influences of the initial angular separation of the satellites on the planet's TTVs, but no significant changes were found.

\section{Conclusions} \label{conclusion} 
In this work, we studied the stability of co-orbital exomoons using the candidates Kepler-1625 b-I and Kepler-1708 b-I as case studies. The proposed exomoons are predicted to be Super-Earth-like satellites. Thus, we opted to work with massive planet-sized satellites as co-orbital companions. We considered bodies with mass and size varying from Mars-like to a body with the same physical attributes of the respective proposed satellite. The latter being chosen so we will have co-orbital satellites with the same mass and size.  

We considered two scenarios, with and without the star as the central body. In the first case, the planet is the central object and the system's dynamics are modeled by the general three-body problem. Adding the star, we increase the complexity and reality of the problem. The gravitational effects of the star will be of great relevance for the Kepler-1625 system since the host planet, in this case, have a semimajor axis smaller than $1$ au. Also, considering the star, we can predict the effects of co-orbital satellites on the planet's TTV, which is not found in the literature yet.

This work aimed to:
\begin{enumerate}
    \item Verify the conditions for stability of co-orbital massive exomoons on the Kepler-1625 b and the Kepler-1708 b systems;
    \item Study the role of libration resonances on the stability of the systems;
    \item Understand the correlation between the amplitude of libration of the co-orbital companion satellite and the perturbations from the primary satellite;
    \item Measure the effects of the system's star over co-orbital exomoons;
    \item Provide planetary TTV profiles for systems with co-orbital exomoons.
\end{enumerate}

In the following, we discuss our results for the systems with and without the star separately.

\subsection{Local system: Planet-Satellite-Co-orbital companion}

As predicted by \citet{Gascheau-1843} and stated later by \citet{Erdi-Sandor-2005}, the vicinity of $L_4$ was indeed stable for most of the different co-orbital companions we tested. However, we find that for less massive satellites (Mars-like bodies), this region can extend for initial angular separations between $48^{\circ}$ and $84^{\circ}$ for the Kepler-1625 system and from $51^{\circ}$ to $79^{\circ}$ for the Kepler-1708 system. The extension of the stable region slowly decreases as we increased the mass and size of the co-orbital companion until we found stable configurations only near $L_4$, as initially expected. Also, we found instability islands for specific values of co-orbital satellites' mass, $M_2 = 5 - 8$ $M_{\oplus}$ and $M_2 = 11 - 12$ $M_{\oplus}$ for the Kepler-1625 system, and $M_2 = 1.5 - 2.0$ $M_{\oplus}$ and $M_2 = 3 - 3.5$ $M_{\oplus}$ for the Kepler-1708 system. For these systems, not even the initial conditions placed at $L_4$ survived.

To understand the causes of the instabilities as a function of the mass of the co-orbital companion, we went back to the results of the restricted three-body problem and drew comparisons between our results and the classical ones.

In \citet{Erdi-Sandor-2005}, the authors mapped the stability of co-orbital systems with different mass parameters under the assumptions of the restricted-three body problem, considering two massive bodies and a particle. They found that librational resonances play a decisive role in raising instabilities depending on the mass parameter of the system. Also, some of these resonances, the 2:1 for example, are so strong that the co-orbital region is unstable for particles, even if the secondary body is in initially circular motion around the primary body.

First of all, we showed that similarly to the restricted case, in our systems, the motion of the co-orbital satellites is described by the superposition of two different motions: a short-period epicyclic motion about the epicentre; and a long-period motion of the epicentre about $L_4$. Thus, we calculate an approximation for the frequencies of these two motions using Eqs. \ref{eq:freq1} and \ref{eq:freq2}, derived for the restricted-three body problem, with the mass parameter $\bar{\mu}$ given by Eq. \ref{eq:mass-parameter}. 

From our analysis, we found that the 2:1 and 5:3 libration resonances may be responsible for the islands of instabilities we detected as we increased the mass of the co-orbital satellites to some specific values. Our results corroborate the findings of \citet{Erdi-Sandor-2005} and show that some characteristics of the restricted three-body problem may be valid for the general case. Also, we searched for the location of the 3:2 librational resonance and found that the stable systems at this location have satellites with low eccentricity. The same behaviour was spotted in the restricted case.

In addition to the librational resonances driven by the mass parameter of the system, we found isolated unstable initial conditions located inside islands of stability ($M_2 = 0.107$ $M_{\oplus}$ and $\theta = 51^{\circ}$, and $M_2 = 15$ $M_{\oplus}$ and $\theta = 56^{\circ}$ for the Kepler-1625 system, and $M_2 = 5$ $M_{\oplus}$ and $\theta = 54-55^{\circ}$ for the Kepler-1708 system). These instabilities appeared as a function of the initial angular separation of the co-orbital pair.

To find the nature of these two unstable conditions, we generated a time series of the evolution of the angular separation of the satellites and applied an FFT analysis to this series to search for the dominant frequencies of the problem for each case. 

For all systems, we found two dominant frequencies. Taking the ratio of the respective frequencies for each system we found some resonances between the motion of the satellites. For the Kepler-1625 system, we found that the instability for the case with $M_2 = 0.107$ $M_{\oplus}$ and $\theta = 51^{\circ}$ was generated by a 5:2 resonance between the motion of the satellites, while for the system with $M_2 = 15$ $M_{\oplus}$ and $\theta = 56^{\circ}$ we found that a 3:2 resonance was the cause of the instability. In these two situations, the resonances increased the eccentricity of the satellites, which ultimately led to a collision between the co-orbital bodies. The same pattern appeared for the Kepler-1708 system, in this case, a 2:1 resonance was responsible for the instabilities when the co-orbital companion was a $M_2 = 5$ $M_{\oplus}$ body initially at $\theta = 54^{\circ}$ and $55^{\circ}$.

Also, we investigated the amplitude of libration of the stable satellites. Here, we gave special attention to the motion of the co-orbital companion, once the movement of the primary satellite in the rotating frame is almost negligible.

The amplitude of libration of the co-orbital satellites is proportional to the mass of the satellite and its initial angular separation. As $\bar{\mu}$ decreases, stable orbits with larger amplitudes of libration become possible \citep{Leleu-etal-2018}. This result was confirmed in our simulations. We found that Mars-like satellites have orbits with larger amplitudes when compared with more massive satellites for the Kepler-1625 system. The surviving co-orbital companions in the Kepler-1708 system presented wider motion amplitude when their masses were $M_2 = 0.5$ $M_{\oplus}$ instead of $M_2 = 0.107$ $M_{\oplus}$, but given the proximity of these values, our conclusions remain unaffected. Horseshoe orbits are not stable for the values of $\bar{\mu}$ considered here. Thus, we did not find this configuration in our results.

On the other hand, we found that co-orbital companions initially farther from $L_4$ are likely to present larger amplitude in their motions. For these cases, the perturbations from the primary satellite and the planet are more pronounced and the amplitude of the tadpole orbit described by the co-orbital companions on the rotating frame increases. If the satellites are angularly closer and suffer a close encounter then the system might become unstable. After reaching instability, the co-orbital satellite may collide with the primary satellite or the planet or even be ejected from the system.

\subsection{Complete system: Star-Planet-Satellite-Co-orbital companion}

The addition of the star proved to be catastrophic for the survival of co-orbital satellites in the Kepler-1625 system. We found that despite the satellites being deep within the Hill radius of the planet, the gravitational influence of the star is still enough to break the co-orbital architectures of the satellites. We found that the most common outcome for the satellites is the collision between them. This is expected given that the satellites initially share the same orbit and experiment close encounters when the initial architecture breaks. In conclusion, massive co-orbital satellites are unlikely on the Kepler-1625 system given the initial conditions we assumed for the planet and satellites.

Our results do not affect previous findings regarding the stability of multiple satellites in the Kepler-1625 system. Here, the initial co-orbital configuration is a major constraint of the problem.

On the other hand, the co-orbital satellites in the Kepler-1708 systems are only marginally disturbed by the addition of the star. This is mainly due to planet-star separation and mass ratio and the initial position of the satellites, inside $5\%$ of the planet's Hill radius.

Once we found stable co-orbital satellites for the Kepler-1708 system, we analyzed the influences of this type of configuration on the planet's transit timing variation. As expected, the amplitude of the TTV increased as the mass of the co-orbital companion increased. For $M_2 = 5$ $M_{\oplus}$, we found an increase of about $5$ minutes on the amplitude of the TTV compared with the case with only one moon. For smaller co-orbital companions, these effects are more subtle. The initial angular position of the co-orbital satellites is not relevant for the TTV produced by the planet.

Finally, the results presented here are dependent on the initial conditions adopted. For the Kepler-1625 system, there are papers showing that the planet-satellite separation can be smaller than the adopted $40$ $R_p$. This consideration alone will place the satellite in a position where the effects of the star will be less relevant, consequently increasing the possibility of finding stable co-orbital satellites. Moreover, for the local systems, we considered that the planet and the satellites are in the same orbital plane, even though the planet is thought to be inclined regarding its star. When adding the star, we simulated the cases with the planet coplanar and inclined ($I_p = 45^{\circ}$), which did not change the scenario of instability for co-orbital satellites. The Kepler-1708 system has even more uncertainties than the previous system. We opted to consider the system's properties presented in the paper that proposed the exomoon candidate \citep{Kipping-etal-2022} and in the study that explored the tidal evolution of the satellite \citep{Tokadjian-Piro-2022a}. All in all, more details about these systems are needed to build more accurate models.  

\section*{Acknowledgements}
RAM dedicates this paper to his late mentor and friend Willy Kley. The authors thank the anonymous referee for the valuable comments and suggestions that significantly improved this manuscript and Muller Lopes for the help with the TTV analysis. This work was possible thanks to the scholarship granted from the Brazilian Federal Agency for Support and Evaluation of Graduate Education (CAPES), in the scope of the Program CAPES-PrInt, process number 88887.310463/2018-00, Mobility number 88887.583324/2020-00 (RAM). RAM, OCW, and DCM thank the financial support from FAPESP (Grant: 2016/24561-0) and CNPq (Grant: 305210/2018-1). This research was supported by resources supplied by the Center for Scientific Computing (NCC/GridUNESP) of the S\~{a}o Paulo State University (UNESP).
\section*{Data Availability}
The data underlying this paper will be shared on reasonable request to the corresponding author.

\section*{ORCID iDs}
R. A. Moraes \orcidicon{0000-0002-4013-8878} \href{https://orcid.org/0000-0002-4013-8878}{https://orcid.org/0000-0002-4013-8878}\\
G. Borderes-Motta \orcidicon{0000-0002-4680-8414} \href{https://orcid.org/0000-0002-4680-8414}{https://orcid.org/0000-0002-4680-8414}\\
O. C. Winter \orcidicon{0000-0002-4901-3289} \href{https://orcid.org/0000-0002-4901-3289}{https://orcid.org/0000-0002-4901-3289}\\
D. C. Mour\~{a}o \orcidicon{0000-0001-9555-8143} \href{https://orcid.org/0000-0001-9555-8143}{https://orcid.org/0000-0001-9555-8143}\\



\bibliographystyle{mnras}
\bibliography{paper}

\bsp	
\label{lastpage}
\end{document}